\documentclass[preprint]{aastex62}
\usepackage{color}

\usepackage{multirow}

\shortauthors{Yamane et al.}

\begin{document}
\title{ALMA observations of supernova remnant N~49 in the LMC:\\
I. Discovery of CO clumps associated with X-ray and radio continuum shells}


\author{Y. Yamane}
\affiliation{Department of Physics, Nagoya University, Furo-cho, Chikusa-ku, Nagoya 464-8601, Japan; yamane.y@a.phys.nagoya-u.ac.jp}

\author{H. Sano}
\affiliation{Department of Physics, Nagoya University, Furo-cho, Chikusa-ku, Nagoya 464-8601, Japan; yamane.y@a.phys.nagoya-u.ac.jp}
\affiliation{Institute for Advanced Research, Nagoya University, Furo-cho, Chikusa-ku, Nagoya 464-8601, Japan; sano@a.phys.nagoya-u.ac.jp}

\author{J. Th. van Loon}
\affiliation{Lennard-Jones Laboratories, Keele University, Staffordshire ST5 5BG, UK}

\author{M. D. Filipovi$\mathrm{\acute{c}}$}
\affiliation{Western Sydney University, Locked Bag 1797, Penrith South DC, NSW 1797, Australia}

\author{K. Fujii}
\affiliation{Department of Astronomy, School of Science, The University of Tokyo, 7-3-1 Hongo, Bunkyo-ku, Tokyo 133-0033, Japan}

\author{K. Tokuda}
\affiliation{Department of Astrophysics, Graduate School of Science, Osaka Prefecture University, 1-1 Gakuen-cho, Naka-ku, Sakai 599-8531, Japan}
\affiliation{National Astronomical Observatory of Japan, Mitaka, Tokyo 181-8588, Japan}

\author{K. Tsuge}
\affiliation{Department of Physics, Nagoya University, Furo-cho, Chikusa-ku, Nagoya 464-8601, Japan; yamane.y@a.phys.nagoya-u.ac.jp}

\author{T. Nagaya}
\affiliation{Department of Physics, Nagoya University, Furo-cho, Chikusa-ku, Nagoya 464-8601, Japan; yamane.y@a.phys.nagoya-u.ac.jp}

\author{S. Yoshiike}
\affiliation{Department of Physics, Nagoya University, Furo-cho, Chikusa-ku, Nagoya 464-8601, Japan; yamane.y@a.phys.nagoya-u.ac.jp}

\author{K. Grieve}
\affiliation{Western Sydney University, Locked Bag 1797, Penrith South DC, NSW 1797, Australia}

\author{F. Voisin}
\affiliation{School of Physical Sciences, University of Adelaide, North Terrace, Adelaide, SA 5005, Australia}

\author{G. Rowell}
\affiliation{School of Physical Sciences, University of Adelaide, North Terrace, Adelaide, SA 5005, Australia}

\author{R. Indebetouw}
\affiliation{Department of Astronomy, University of Virginia, Charlottesville, VA 22904, USA}
\affiliation{National Radio Astronomy Observatory, 520 Edgemont Rd, Charlottesville, VA 22903, USA}

\author{M. Laki{\'c}evi{\'c}}
\affiliation{Astronomska opservatorija Beograd; Volgina 7, 11000 Beograd, Serbia}

\author{T. Temim}
\affiliation{Space Telescope Science Institute, 3700 San Martin Drive, Baltimore, MD 21218, USA}

\author{L. Staveley-Smith}
\affiliation{International Centre for Radio Astronomy Research (ICRAR), University of Western Australia, 35 Stirling Highway, Crawley, WA 6009, Australia}
\affiliation{ ARC Centre of Excellence for All-Sky Astrophysics (CAASTRO), Building A28, School of Physics, The University of Sydney, NSW 2006, Australia}

\author{J. Rho}
\affiliation{SETI Institute, 189 Bernardo Ave, Mountain View, CA 94043}

\author{K. S. Long}
\affiliation{Space Telescope Science Institute, 3700 San Martin Drive, Baltimore, MD 21218, USA}

\author{S. Park}
\affiliation{Department of Physics, University of Texas at Arlington, Box 19059, Arlington, TX 76019, USA}

\author{J. Seok}
\affiliation{Department of Physics and Astronomy, University of Missouri, Columbia, MO 65211, USA}
\affiliation{Key Laboratory of Optical Astronomy, National Astronomical Observatories, Chinese Academy of Sciences, Beijing 100012, China}

\author{N. Mizuno}
\affiliation{National Astronomical Observatory of Japan, Mitaka, Tokyo 181-8588, Japan}

\author{ A. Kawamura}
\affiliation{Department of Physics, Nagoya University, Furo-cho, Chikusa-ku, Nagoya 464-8601, Japan; yamane.y@a.phys.nagoya-u.ac.jp}

\author{ T. Onishi}
\affiliation{Department of Astrophysics, Graduate School of Science, Osaka Prefecture University, 1-1 Gakuen-cho, Naka-ku, Sakai 599-8531, Japan}

\author{ T. Inoue}
\affiliation{Department of Physics, Nagoya University, Furo-cho, Chikusa-ku, Nagoya 464-8601, Japan; yamane.y@a.phys.nagoya-u.ac.jp}

\author{S. Inutsuka}
\affiliation{Department of Physics, Nagoya University, Furo-cho, Chikusa-ku, Nagoya 464-8601, Japan; yamane.y@a.phys.nagoya-u.ac.jp}

\author{K. Tachihara}
\affiliation{Department of Physics, Nagoya University, Furo-cho, Chikusa-ku, Nagoya 464-8601, Japan; yamane.y@a.phys.nagoya-u.ac.jp}

\author{Y. Fukui}
\affiliation{Department of Physics, Nagoya University, Furo-cho, Chikusa-ku, Nagoya 464-8601, Japan; yamane.y@a.phys.nagoya-u.ac.jp}
\affiliation{Institute for Advanced Research, Nagoya University, Furo-cho, Chikusa-ku, Nagoya 464-8601, Japan; sano@a.phys.nagoya-u.ac.jp}

\begin{abstract}
N49 (LHA~120-N49) is a bright X-ray supernova remnant (SNR) in the Large Magellanic Cloud. We present new $^{12}$CO($J$ = 1--0, 3--2), H{\sc i}, and 1.4 GHz radio-continuum observations of the SNR N49 using Mopra, ASTE, ALMA, and ATCA. We have newly identified three H{\sc i} clouds using ATCA with an angular resolution of $\sim20''$: one associated with the SNR and the others located in front of the SNR. Both the CO and H{\sc i} clouds in the velocity range from 280--291 km s$^{-1}$ are spatially correlated with both the soft X-rays (0.2--1.2 keV) and the hard X-rays (2.0--7.0 keV) of N49 on a $\sim$10 pc scale. CO 3--2/1--0 intensity ratios indicate higher values of the CO cloud toward the SNR shell with an angular resolution of $\sim45''$, and thus a strong interaction was suggested. Using the ALMA, we have spatially resolved CO clumps embedded within or along the southeastern rim of N49 with an angular resolution of $\sim3''$. Three of the CO clumps are rim-brightened on a 0.7--2 pc scale in both hard X-rays and the radio continuum$:$ this provides further evidence for dynamical interactions between the CO clumps and the SNR shock wave. The enhancement of the radio synchrotron radiation can be understood in terms of magnetic-field amplification around the CO clumps via a shock-cloud interaction. We also present a possible scenario in which the recombining plasma that dominates the hard X-rays from N49 was formed via thermal conduction between the SNR shock waves and the cold$/$dense molecular clumps.
\end{abstract}

\vspace*{2cm}
\section{Introduction}
In young supernova remnants (SNRs), interactions between the SNR shock waves and the interstellar medium have received much attention as an important process for the origin of cosmic rays and high-energy radiation. The shock-cloud interaction excites turbulence that intensifies the magnetic-field up to $\sim$1 mG around the shocked gas clumps \citep[e.g.,][]{2007Natur.449..576U,2009ApJ...695..825I,2012ApJ...744...71I}. In regions where synchrotron radiation is enhanced, cosmic-ray electrons are efficiently accelerated to 1 TeV or higher  \citep[e.g.,][]{2015ApJ...799..175S, 2016scir.book.....S}. On the other hand, \cite{2017JHEAp..15....1S} found that the thermal X-ray flux is proportional to the column density of neutral gas associated with the Galactic SNR RCW~86. Interstellar gas interacting with SNRs also provides a target for cosmic rays, enabling the production of high-energy gamma-rays via neutral-pion decay {\citep[e.g.,][]{1994JPhG...20..477N}}. The spatial correspondence between the gamma-rays and the interstellar gas provides conclusive proof of cosmic-ray acceleration in {a SNR} \citep[][]{2003PASJ...55L..61F, 2012ApJ...746...82F, 2017ApJ...850...71F, 2013ApJ...768..179Y, 2014ApJ...788...94F}.

In studying the shock-cloud interactions in Galactic SNRs, it is generally difficult to identify the interstellar gas specifically associated with the shock waves because of huge contamination along the line of sight. There are also large uncertainties in determining the distances to Galactic SNRs, which cause large errors in the determination of physical properties, e.g., the age and size of the SNR, the gas density and mass, and the absolute luminosity of the radiation. We have therefore focused on the SNRs in the Large Magellanic Cloud (LMC) to avoid these problems. The LMC is the nearest star-forming galaxy, in which 59 SNRs have been cataloged in X-rays and some 20 other objects as SNR candidates \cite[e.g.,][]{2016A&A...585A.162M, 2017ApJS..230....2B}. Since the LMC is an almost face-on galaxy \cite[$\sim$20--30 degree; e.g.,][]{2010A&A...520A..24S} relative to our line of sight and since its distance has been determined to be 50 $\pm$ 1.3 kpc\footnote{The inclination of the LMC may cause an error in distance up to $\sim$10$\%$ \cite{2010A&A...520A..24S}. In this paper, we adopt the distance 50 kpc to compare physical properties of other SNRs in the LMC.} \citep{2013Natur.495...76P}, we can easily identify the interstellar gas associated with the SNR, suffering very little contamination along the line of sight. In addition, the redshift of the LMC also helps to separate any foreground {gas local to the sun} from the LMC gas. The LMC is therefore the good laboratory for studying the interactions between SNR shock waves and the interstellar gas. 

N49 (LHA 120-N49; also known as DEM L190 and MCSNR J0526--6605) is a bright X-ray SNR. The SNR is located in the northern part of the LMC, and its coordinates are ($\alpha_{\rm J2000}, \delta_{\rm J2000}$) $=$ ($5^{\mathrm h}26^{\mathrm m} 01^{\mathrm s}, -66^{\mathrm d} 05^{\mathrm m} 06^{\mathrm s}$) \citep{1956ApJS....2..315H}. Its age is $\sim$5000 yr \citep{2012ApJ...748..117P}. N49 has been observed at various wavelengths, including radio, optical, infrared, ultraviolet, and X-rays \citep[e.g.,][]{1998AJ....115.1057D, 2007AJ....134.2308B, 2010A&A...518L.139O, 2003ApJ...586..210P, 2004AJ....128.1615S}. At the far-infrared wavelength, \cite{2010AJ....139...68V} indicated that N49 shows spectacular [O{\sc i}] 63-micron line emission which comes from associated gas and the local interstellar medium.

\cite{2015ApJ...799...50L} presented that the dust within LMC supernova remnants was most likely collisionally heated and sputtered. The optical and infrared wavelengths {of N49} show an asymmetric morphology, with bright filaments in the southwest, and emission from the northwest is lacking. In addition, the radio continuum exhibits clear shell structure, and the X-ray image is bright inside the radio shell  \cite[e.g.,][]{1995ApJ...448..623D, 1992ApJ...394..158V}. 
The brightness of the radio continuum and the X-rays increases in the southeast, and the intensity ratio of the southeast to the northwest portion of the shell is $\sim$9 for the X-rays and $\sim$3 for the radio continuum \citep{1998AJ....115.1057D}.

Previous X-ray studies of N49 have been performed using $ROSAT$, $ASCA$, $Chandra$, $Suzaku$, and $XMM$--$Newton$ \cite[e.g.,][]{1996ApJ...470..513M, 1998ApJ...505..732H, 2003ApJ...586..210P, 2012ApJ...748..117P, 2014ApJ...785L..27Y, 2016A&A...585A.162M}. \cite{1982ApJ...255L..45C} and \cite{1994Natur.368..432R} found the X-ray counterpart of the soft gamma-ray repeater (SGR) 0525$-$66 to be located inside the SNR shell. Using $Chandra$, \cite{2003ApJ...585..948K} determined its coordinates to be specified ($\alpha_{\rm J2000}, \delta_{\rm J2000}$) $=$ ($5^{\mathrm h}26^{\mathrm m} 0\fs9, -66^{\mathrm d} 04^{\mathrm m} 36^{\mathrm s}$) with an error of 0\farcs6. \cite{2015ApJ...808...77U} reported the discovery of recombining plasma in N49 using detailed spectroscopy from $Suzaku$. The particular characteristic of recombining plasma is that the ionization temperature is higher than the electron temperature, thus recombination dominates ionization. This type of plasma has a characteristic spectrum mainly in the hard X-ray band ($\varepsilon$ $>$ 2 keV) and has been detected in more than 10 mixed-morphology SNRs \cite[see][and references therein]{2015ApJ...808...77U}.

Whether the explosion that produced N49 was a core-collapse or a Type Ia supernova has been debated ever since N49 was discovered \cite[e.g.,][]{2012ApJ...748..117P, 2017ApJS..230....2B}. \cite{2017ApJS..230....2B} argued that the radio luminosity of Type Ia SNRs are significantly weaker {than} core-collapse SNRs and N49 is among the brightest SNRs (with 1.66 Jy flux density at 1 GHz) in the LMC making it very unlikely Type Ia explosion. {The} physical association between the SGR 0525--66 and the SNR {may also support the core-collapse origin of N49} \cite[e.g.,][]{2001ApJ...559..963G, 2004ApJ...609L..13K, 2009ApJ...700..727B}. \cite{2004ApJ...609L..13K} argued that the SGR is related to the OB association LH 53, which lies at a distance $\sim$30 pc from the SGR, and that the progenitor of N49 was a B-type star. \cite{1985MNRAS.212..799S} pointed out that a B-star progenitor for N49 would have formed {H{\sc ii} region}. {On the other hand, \cite{2012ApJ...748..117P} showed that the Si/S ejecta mass ratio in N49 may not clearly discriminate between core-collapse and Type Ia origins.}


N49 also is known to be {an} SNR that is interacting with a dense cloud. \cite{1997ApJ...480..607B} discovered the dense molecular cloud along the southeastern rim of N49 using the Swedish-ESO Submillimeter Telescope (SEST). They observed the $^{12}$CO($J$ = 2--1) line utilizing the position-switching technique. The molecular cloud has a radial velocity of $\sim$286 km s$^{-1}$ and a radius of $\sim$7.2 pc, and it appears to be physically connected with the bright radio continuum and the X-ray shell of the SNR. \cite{2003ApJ...586..210P} mentioned that the soft X-ray emission is enhanced near the molecular cloud. In previous studies, however, the clumpy structure of the molecular cloud could not be detected because of the poor angular resolution of $\sim$23$\arcsec$ ($\sim$5.5 pc at the LMC) and under-sampled observations. The distribution and kinematics of the atomic gas traced by the H{\sc i} are also unknown. Moreover, detailed comparisons among the CO, H{\sc i}, X-rays, and radio continuum have not previously been performed. 

In this study, we present new $^{12}$CO($J$ = 1--0, 3--2), H{\sc i}, and 1.4 GHz radio-continuum datasets for the SNR N49 using the Mopra radio telescope, the Atacama Submillimeter Telescope Experiment (ASTE), the Atacama Large Millimeter/Submillimeter Array (ALMA), and the Australia Telescope Compact Array (ATCA). Using these data, we investigate the physical relations among the molecular/atomic clouds, radio continuum, and X-rays to understand the origins of the synchrotron radio emission and the recombining plasma. Owing to the high-spatial resolution $\sim$3$\arcsec$ of ALMA data, we can detect molecular clumps with $\sim$0.7 pc resolution. Section \ref{sec:obs} describes the observations and reductions of the CO, H{\sc i}, and X-ray datasets. Section \ref{sec:res} comprises five subsections: Subsection \ref{subsec:ldis} and \ref{subsec:azimuth} describe the large-scale CO, H{\sc i}, and X-ray distributions of N49; Subsection \ref{subsec:ratio} gives an intensity ratio of CO $J$ = 3--2 / 1--0; Subsections \ref{subsec:ddis} and \ref{subsec:dcom} present detailed CO distributions for N49 and compares them with the X-rays. Sections \ref{sec:dis} and \ref{sec:sum}, respectively, discuss our results and {summarizes} our conclusions.

\section{Observations and data reduction} \label{sec:obs}
\subsection{CO}\label{subsec:CO}
\subsubsection{Mopra}\label{subsub:Mopra}

The $^{12}$CO($J$ = 1--0) emission line data {were} taken using the Mopra 22-m radio telescope of the Australia Telescope National Facility (ATNF) from 2014 May to July (PI: Kosuke Fujii). The on-the-fly (OTF) mode with Nyquist sampling was used with mapping areas of $3' \times 3'$. The typical system temperature was 600--1000 K in the single-side band (SSB). The Mopra Spectrometer (MOPS) as the backend system offered 4096 channels across 137.5 MHz, which correspond to a velocity resolution of 0.088 km s$^{-1}$  per channel. The velocity coverage is 360 km s$^{-1}$ in the zoom mode at 115 GHz. After convolution with a 2D Gaussian kernel, the final beam size of the datasets was $\sim$45$\arcsec$ ($\sim$11 pc at the LMC). The pointing accuracy was checked for every 2 hours to hold within  an offset of 7$''$. The intensity calibration was applied by observing Orion-KL assuming the peak antenna temperature of $T_{\mathrm{A}}^{\ast} = 45.9$ K divided by the extended beam efficiency of $\eta_\mathrm{XB} \sim 0.46$ to be consistent with corrected peak CO brightness temperature of $T_\mathrm{XB} \sim 100$ K \citep{2005PASA...22...62L}. By using the root-mean-square weighting, we combined our data with the archived datasets from the Magellanic Mopra Assessment \citep[MAGMA; ][]{2011ApJS..197...16W, 2017ApJ...850..139W} to improve the noise fluctuation. The final noise fluctuation was $\sim$0.1 K at a velocity resolution of 1 km s$^{-1}$.

\subsubsection{ASTE}\label{subsub:ASTE}
The $^{12}$CO($J$ = 3--2) observations {were} performed by using ASTE \citep{2004SPIE.5489..763E} during August 2014. The OTF mode with Nyquist sampling was used with mapping areas of $3' \times 3'$. The frontend was ``CATS 345'' which is a 2SB SIS mixer receiver \citep{2008stt..conf..281I} and the typical system temperature was 250--350 K in the SSB. The digital spectrometer, ``MAC'' \citep{2000SPIE.4015...86S} as the backend offered 1024 channels across 128 MHz, which correspond to a velocity resolution of $\sim$ 0.11 km s$^{-1}$  per channel and the velocity coverage is $\sim$ 111 km s$^{-1}$ at 350 GHz. The pointing accuracy was checked for every 1 hour to hold within an offset of 6$''$. As a standard calibrator source, N159W was observed with the peak antenna temperature of $T_{\mathrm{A}}^{\ast} = 7.99$ K and the intensity divided the scaling factor of $\eta \sim 0.57$ to convert the main beam temperature $(T_\mathrm{MB})$ scale \citep{2011AJ....141...73M}. The beam size was 45$''$ that was smoothed to match the FWHM of the Mopra $^{12}$CO($J$ = 1--0) data. The final noise fluctuation was $\sim0.02$ K at the velocity resolution of 0.53 km s$^{-1}$.

\subsubsection{ALMA}\label{subsub:ALMA}
We carried out ALMA Cycle 3, Band 3 (86--116 GHz) observations (P.I., Jacco Th. van Loon) of $^{12}$CO($J$ = 1--0) emission line, and the 1.3-mm continuum toward N49, centered on the position of ($\alpha_{\rm J2000}, \delta_{\rm J2000}$) $=$ ($5^{\mathrm h} 26^{\mathrm m} 4\fs00, -66^{\mathrm d} 05^{\mathrm m} 20\fs0$), using 49 antennas of the ALMA 12-m array. The observations were carried out on January 12 and 18, 2016, through three spectral windows. The target molecular lines were $^{12}$CO($J$ = 1--0) with a bandwidth of 234.4 MHz (141.1 kHz $\times$ 3840 channels). We used a spectral window with a bandwidth of 1875 MHz (31.25 MHz $\times$ 128 channels) for observations of the continuum emission. The baselines ranged from 13.0 to 282.7 m, corresponding to $u$--$v$ distances from 5.0 to 108.6 k$\lambda$. The calibration of the complex gains was carried out using observations of J0635--7516, phase calibration on J0516--6207, and flux calibration on J0519--4546. 
For imaging process from the visibility data, we used the Common Astronomy Software Applications package \citep [CASA, version 4.7.2;][]{2007ASPC..376..127M}. We applied the natural weighting to recover the weak extended emission. We also applied the multi-scale CLEAN algorithm \cite{2008ISTSP...2..793C} implemented in CASA. The multi-scale CLEAN approach is useful to better restore the extended emission and reduce negative structures.
The synthesized beam size was $\sim$$3\farcs2$ $\times$ $2\farcs2$ ($\sim$0.8 $\times$ 0.5 pc at 50 kpc) with a position angle of 66.1 degrees, and the rms noise was $\sim$8 mJy beam$^{-1}$ ($\sim$0.07 K), with a velocity resolution of 0.4 km s$^{-1}$.

To derive the missing flux, we compared the intensity of the single-dish (Mopra) data integrated over the velocity from 282.9 to 288.7 km s$^{-1}$ and our re-processed ALMA data smoothed to the single-dish resolution $\sim45''$. As a result, the ALMA data is $\sim3\%$ smaller than the single-dish data and thus the missing flux is considered to be negligible.

\subsection{H{\sc i} and radio continuum}\label{subsec:HI}
The H{\sc i} 21 cm data are from K. Fujii et al. (2018, in preparation), and they were obtained with ATCA and the Parkes radio telescope (C2908, PI: Kosuke Fujii). The combined H{\sc i} image has an angular resolution of 24$\farcs$75 $\times$ 20$\farcs$48 with a position angle of $-$35 degrees, and the noise fluctuations of the final data cube were $\sim$2.7 K for a velocity resolution of 0.4 km s$^{-1}$. As for the continuum data, we used same ATCA observations (C2908, PI: Kosuke Fujii) but in continuum (mosaic) mode with all three available arrays (1.5A, 1.5B, and 1.5D) and standard primary (1934$-$638) and secondary (0407$-$658) calibrator.  For all our ATCA data reduction, we used {Miriad} software \citep{1995ASPC...77..433S}. The best radio continuum images were obtained using neutral weighting scheme of ${\rm robust} = 0$ and self-calibration. The combined radio-continuum image has an angular resolution of 5$\farcs$33 $\times$ 3$\farcs$87 with a position angle of $-$13.7 degrees, and the noise fluctuations were $\sim$70 $\mu$Jy beam$^{-1}$.

\subsection{X-rays}\label{subsec:X-rays}
We used archived X-ray data obtained from the $Chandra$ Data Archive. Observations of N49 were conducted with the Advanced CCD Imaging Spectrometer S-array (ACIS-S3) on July 18--September 19, 2009, during AO10 (Obs IDs 10123, 10806, 10807, and 10808; PI: Sangwook Park). The total exposure time was 108 ks. We reduced these observations using the CIAO v4.9 software package \citep{2006SPIE.6270E..1VF} with CALDB v4.7.6. Each dataset was processed using the {\bf chandra\_repro} script. To create the combined energy-filtered and exposure-corrected images, we used the {\bf merge\_obs} script. The pixel size of the created data is 0$\farcs$984.

\section{Results} \label{sec:res}
\subsection{The large-scale CO, H{\sc i}, RC (radio continuum) and X-ray distributions of N49}\label{subsec:ldis}
We first present the large-scale distributions of the SNR N49 obtained with Mopra $^{12}$CO($J$ = 1--0), ATCA \& Parkes H{\sc i}/RC, and the $Chandra$ X-ray datasets. Figure \ref{fig1} shows a three-color X-ray image of N49 and white contours are RC (1.4~GHz). Both the soft X-rays (red: 0.5--1.2 keV) and the hard X-rays (blue: 2.0--7.0 keV) are enhanced in the southeastern direction. The SGR 0525--66 is located in the northeast of the SNR at ($\alpha_{\rm J2000}, \delta_{\rm J2000}$) $\sim$ ($5^{\mathrm h} 26^{\mathrm m} 00\fs7, -66^{\mathrm d} 04^{\mathrm m} 35\fs0$), and the ejecta bullet appears on the western rim at ($\alpha_{\rm J2000}, \delta_{\rm J2000}$) $\sim$ ($5^{\mathrm h} 26^{\mathrm m} 52\fs5, -66^{\mathrm d} 05^{\mathrm m} 14\fs6$){, which were previously mentioned by \cite{2003ApJ...586..210P, 2012ApJ...748..117P}}. These spots are also bright in hard X-rays. We find the X-ray bullet in the RC image as well. We also note that the X-ray peak in the southeastern shell is significantly offset from the radio continuum peak.

Figures \ref{fig3}(a), \ref{fig3}(b), and \ref{fig3}(c) show the H{\sc i} distributions at velocity ranges of {$V_{\rm LSR}$ $=$ 280--291 km s$^{-1}$ (hereafter refer as the blue-velocity), $V_{\rm LSR}$ $=$ 291--298 km s$^{-1}$ (hereafter refer as the middle-velocity), and $V_{\rm LSR}$ $=$ 298--306 km s$^{-1}$ (hereafter refer as the red-velocity), respectively.} The blue-velocity H{\sc i} cloud is bright along the southeastern edge of the SNR. {For the middle- and red-velocity H{\sc i} clouds, the H{\sc i} brightness temperature is decreased toward the southeastern rim of the SNR, where the radio continuum and X-rays are bright. Hereafter {we} refer to the H{\sc i} intensity depression as ``dip-like structure.'' The distribution of these gases tends to encircle the shell-like structure.} {The morphology of the red and middle gas and its possible association with the SNR is discussed more in section \ref{subsec:ass} below.}

Figures \ref{fig3}(d), \ref{fig3}(e), and \ref{fig3}(f) show the integrated intensity maps for the Mopra $^{12}$CO($J$ = 1--0) {at the blue, middle, and red velocities, respectively.} We detected CO clouds at the blue- and red-velocities{, whereas at the middle-velocity, there is not significant CO emission}. At the blue-velocity, we detected  two molecular clouds. One extends toward the northeast of the SNR, and the other lies near the southeastern rim of the SNR and has a diameter of $\sim$7 pc{, the latter of which is thought to be associated with N49 by \cite{1997ApJ...480..607B}.} {At the red-velocity, the cloud that is $\sim50$ pc in length extends over the SNR, while there is no spatial correlation with the radio continuum shell.}

Figure \ref{fig2} shows the spatially averaged spectra of H{\sc i} and CO. Over the entire region shown in Figure \ref{fig3}, the averaged H{\sc i} spectrum (red line) has two peaks, velocities of $\sim$286 and $\sim$295 km s$^{-1}$, and in the rectangular region outlined in Figure \ref{fig3}(a), the averaged H{\sc i} spectrum (black line) has three peaks, at velocities of $\sim$286, $\sim$295, and $\sim$302 km s$^{-1}${, representing the blue, middle, and red-velocities defined as Figure \ref{fig3}.} The averaged CO spectrum has a peak intensity of 0.2 K at a velocity of $\sim$286 km s$^{-1}$.


Figure \ref{fig4} shows position-velocity diagrams for H{\sc i} and CO. We find cavity-like H{\sc i} structures in the velocity ranges from 286 to 294 km s$^{-1}$ and from 297 to 303 km s$^{-1}$, which have similar diameters to the SNR in declination. The CO cloud also appears in the blue-velocity at $\sim$286 km s$^{-1}$. We hereafter focus on the CO and H{\sc i} clouds at the blue-velocity, which are thought to be associated with the SNR N49 by \cite{1997ApJ...480..607B}.

\subsection{{Azimuth profiles of CO, H{\sc i}, and X-rays}}\label{subsec:azimuth}
Figures \ref{fig5}(a) and \ref{fig5}(b) display the same CO and H{\sc i} images in the blue-velocity as shown in Figures \ref{fig3}(a) and \ref{fig3}(d), but the superposed contours here indicate {broad-band} X-rays {in the energy band of 0.5--7.0 keV}. To evaluate the relation between the {X-rays} and CO/H{\sc i} clouds, we examined their azimuthal distributions. First, we defined the shell boundary based on visual inspection. The shell is centered at  ($\alpha_{\rm J2000}, \delta_{\rm J2000}$) $=$ ($5^{\mathrm h} 25^{\mathrm m} 59\fs5, -66^{\mathrm d} 04^{\mathrm m} 59\fs2$) [see the red dashed lines in Figures \ref{fig5}(a) and \ref{fig5}(b)]. The southeastern direction that is bright in both CO and X-rays is taken as the origin of the azimuthal angle, which is measured counterclockwise. Figures \ref{fig5}(c) and \ref{fig5}(d) show the azimuthal distributions of the {CO/H{\sc i} emission, the soft X-rays (0.5--1.2 keV), and hard X-rays (2.0--7.0 keV)} between the elliptical rings in Figures \ref{fig5}(a) and \ref{fig5}(b). {The regions that are bright at X-rays} from SGR 0525--66 were excluded from this analysis. The CO distribution shows good spatial correspondence with {both the soft and hard} X-ray{s}. On the other hand, the spatial correspondence between the H{\sc i} and X-rays is not as good as that between the CO and X-rays{, indicating that distribution of CO is possibly more important than that of H{\sc i} to produce the X-rays}. The {linear Pearson} correlation coefficients {are} {$\sim$0.95 for the CO and soft X-rays; $\sim$0.83 for the CO and hard X-rays; $\sim$0.79 for the H{\sc i} and soft X-rays; $\sim$0.65 for the H{\sc i} and hard X-rays.}

\subsection{{CO 3--2/1--0 ratio map}}\label{subsec:ratio}
{Figure \ref{ratio} is the distribution of the line intensity ratio between ASTE $^{12}$CO$(J$ = 3--2) and Mopra $^{12}$CO$(J$ = 1--0) at the blue velocity. The ratio reflects the rotational excitation states of the molecular clouds, and hence the {high-intensity} ratio indicates that the cloud temperature and/or density tend to become high. We find the region with {high-intensity} ratios $\sim 0.6$ along the southeastern rim of N49, where the X-rays and radio continuum are enhanced. The intensity ratio lowers as {it} separates from the southeastern-rim of the SNR. On the other hand, the CO cloud in the northeast of the SNR shows a {low-intensity} ratio $\sim0.2$ or less.}

\subsection{{Detailed CO distributions with ALMA}}\label{subsec:ddis}
Figure \ref{fig6} shows ALMA $^{12}$CO($J$ = 1--0) channel maps overlaid on the hard X-rays toward the SNR N49. We identified CO clumps in the observed region, which we chose according to the following criteria:

\begin{enumerate}
 \item {In the CO intensity channel maps integrated over the velocity range from 281.4 to 288.6 km s$^{-1}$ every 1.2 km s$^{-1}$, a position of peak intensity above 0.48 K km s$^{-1}$ [equal to the 5$\sigma$ level of the ALMA $^{12}$CO($J$ = 1--0) data] is defined as an individual CO peak.}
 \item {When the CO spectrum toward the peak has a single--peaked profile and the distance between the peak positions of the clumps with connective velocity channel {are} less than 3\farcs2 [equal to ALMA $^{12}$CO($J$ = 1--0) beam size], these clumps are considered as a single clump. As a result, we identified 21 CO clumps and confirmed that all clumps have single-peaked spectra.}
\item {The peak radiation temperature $T_\mathrm{R}^\ast $, the full-width at half maximum (FWHM) line width $\Delta V$, and the velocity at the peak intensity $V_{\rm{peak}}$ are derived from a single Gaussian fitting for CO spectra obtained at the peak positions of the CO clumps.}
\item {The total surface area of a clump is defined as the region surrounded by the contour at half the maximum integrated intensity in the integrated intensity map whose velocity integration range is FWHM line width.}
 \item When two or more peaks are contained in one area enclosed by the same contour, they are defined as independent peaks divided at the minimum {in the above intensity map.}
\item {Cloud size is defined as ($A$/$\pi$)$^{0.5}$ $\times 2 $, where $A$ is the total surface area defined by {procedure} 4.}
\end{enumerate}

We summarize the physical properties of the CO clumps in Table \ref{tab1}. The peak radiation temperatures of these clumps are $\sim$1--12 K. Almost all of the clumps have velocities in the range from 283.6 to {287.4} km s$^{-1}$, and the FWHM line width is {0.7--2.0} km s$^{-1}$. The velocity range spanned by the CO clumps corresponds to that of the Mopra CO cloud at the blue-velocity [see Figure \ref{fig3}(d)]. {The typical CO spectra are shown in Figure \ref{spec}. The CO clump P is far from the shock-front and is well fitted by the Gaussian function because of no shock disturbance. Although the CO clumps S {and O are} embedded within the shell, the CO spectra show slight offsets from the Gaussian function at $V_\mathrm{LSR} \sim 287.5$ km s$^{-1}$ for clump O, and at $V_\mathrm{LSR} \sim 284.5$ km s$^{-1}$ for clump S. Since the offset has a small velocity width $\sim1$ km s$^{-1}$, we conclude that t}here is no clear sign of line broadening in the CO spectra. The sizes of the CO clumps are {0.9}--2.0 pc. We derived the masses of the clumps using two different methods. One gives the virial mass, $M_{\rm vir}$, obtained from the virial theorem and the following:

\begin{equation}
  M_{\mathrm {vir}} = 210 R {(\Delta V)}^2 M_\sun,
\label{equ:1}
\end{equation}        
where $R$ and {$\Delta V$} are the radius in unit of pc and the FWHM line width in unit of km s$^{-1}$ of CO clumps, respectively. The other gives the CO-derived mass, $M_{\rm CO}$, obtained from the following equation:

\begin{equation}
  M_{\rm CO} = m_{\rm H} \mu \sum_{i}[D^2 \Omega N_i({\mathrm H}_2)],  
\label{equ:2}
\end{equation}
where $m_{\rm H}$ is the mass of atomic hydrogen, $\mu$ is the mean molecular weight relative to atomic hydrogen, $D$ is the distance to the source in unit of cm, $\Omega$ is the solid angle subtended by a unit grid spacing of a square pixel (0$\farcs$5 $\times$ 0$\farcs$5), and $N_i({\rm H}_2)$ is the molecular hydrogen column density for each pixel in unit of cm$^{-2}$. We adopt $\mu$ = 2.38 to take into account the $\sim$36 \% abundance by mass of helium relative to atomic hydrogen. We also used the following relationship between the molecular hydrogen column density $N$(H$_2$) and the $^{12}$CO($J$ = 1--0) integrated intensity $W$(CO) from \cite{2008ApJS..178...56F}:

\begin{equation}
  N({\rm H}_2) = 7.0 \times 10^{20} W({\rm CO}) ({\rm cm}^{-2}).
\label{equ:3}
\end{equation}
where the units for $W$(CO) are K km s$^{-1}$.

\subsection{Detailed comparison with X-rays}\label{subsec:dcom}
Figure \ref{fig7} shows integrated intensity maps of the ALMA CO in the blue-velocity overlaid on hard X-rays $Chandra$ [Figure \ref{fig7}(a)], soft X-rays [Figure \ref{fig7}(b)], and radio continuum [Figure \ref{fig7}(c)]. The spatial distribution of hard X-rays is roughly consistent with that of the soft X-rays on a 5 pc scale, and both the hard and soft X-rays are bright in the southeastern part of the SNR. There are three hard-X-ray peaks in the southeast of the SNR at ($\alpha_{\rm J2000}, \delta_{\rm J2000}$) $\sim$ ($5^{\mathrm h}$ $26^{\mathrm m}$ 04\fs3, $-66^{\mathrm d}$ $04^{\mathrm m}$ 53\fs6), $\sim$ ($5^{\mathrm h}$ $26^{\mathrm m}$ 04\fs1, $-66^{\mathrm d}$ $05^{\mathrm m}$ 15\fs5), and $\sim$ ($5^{\mathrm h}$ $26^{\mathrm m}$ 01\fs3, $-66^{\mathrm d}$ $05^{\mathrm m}$ 21\fs2). Each X-ray peak appears to be associated with a CO peak on a pc scale. The peak intensity of {soft X-rays,  at  ($\alpha_{\rm J2000}, \delta_{\rm J2000}$) $\sim$ ($5^{\mathrm h}$ $26^{\mathrm m}$ 04\fs7, $-66^{\mathrm d}$ $05^{\mathrm m}$ 05\fs6) and ($\alpha_{\rm J2000}, \delta_{\rm J2000}$) $\sim$ ($5^{\mathrm h}$ $26^{\mathrm m}$ 01\fs9, $-66^{\mathrm d}$ $05^{\mathrm m}$ 26\fs2), and} radio continuum, at  ($\alpha_{\rm J2000}, \delta_{\rm J2000}$) $\sim$ ($5^{\mathrm h}$ $26^{\mathrm m}$ 04\fs7, $-66^{\mathrm d}$ $05^{\mathrm m}$ 05\fs6), {are} also located near the CO clumps. Hereafter, we focus on the CO clumps labeled J, K, and L that are rim-brightened in X-rays. 

Figure \ref{fig8} shows the ALMA CO channel map that enlarges the dashed region from Figure \ref{fig7}(a). The contours indicate hard X-rays in (a), soft X-rays in (b) and the radio continuum in (c). The distribution of hard X-rays is clearly different from that of soft X-rays on a sub-pc scale. The hard X-rays seem to be enhanced around CO clumps J, K, and L. 
{Around clump{s J and L}, the soft X-rays also seem to be enhanced to the different direction from hard X-rays.} The radio-continuum peak is located between the CO clumps {K and L}.

Figure \ref{fig9} shows radial profiles that compare the spatial distributions of CO clumps J, K, and L with those of the hard/soft X-rays and the radio continuum. {We investigated selected rectangular regions to the direction perpendicular to the shock front. The hard X-rays near CO clumps J and K are bright for only half of the CO clumps.} {The spatial separation between the hard X-ray peak and CO clumps J and K} is $\sim$2 pc. On the other hand, the hard X-rays near CO clump {L} are enhanced so as to surround the CO clump, indicating the clump embedded within the shock wave. We therefore create the radial profile, and {the separation between the hard X-ray peak and CO clump L} is $\sim$0.6 pc. On the other hand, the separation between the soft X-ray peak and CO clumps J and K is less than 0.5 pc, which is shorter than the separation between the hard X-ray peak and the CO peaks. 

Figure \ref{fig10} shows detailed comparisons between the distributions of soft/hard X-rays with that of CO clump {K}. In Figure \ref{fig10}(c), the intensity peak of the hard X-rays shows a separation of $\sim$2 pc from that of the CO clump, whereas the soft X-rays show a separation of $\sim$3 pc from the CO clump. In strip B of Figure \ref{fig10}(a), the hard X-ray intensity abruptly increases at (Offset X, Offset Y) $\sim$ ($-$1$\arcsec$, 1$\farcs$5), whereas the soft X-rays still exhibit a {low-intensity} in the same region [see Figure \ref{fig10}(b)]. We also find that the molecular clumps are distributed only over a region of $\ga$ 4$\arcsec$ in Figures 10(a) and 10(b), and the hard X-ray intensity at $\sim$4$\arcsec$ in Figures 10(a) is 40 \% higher than the soft-X-ray intensity at $\sim$4$\arcsec$ in Figure \ref{fig10}(b).

\section{Discussion} \label{sec:dis}
\subsection{{Atomic and molecular clouds associated with the SNR N49}}\label{subsec:ass}
The H{\sc i} and CO clouds toward N49 are concentrated in the three red-, middle-, and blue- velocity components (see Figures \ref{fig2} and \ref{fig3}). In this section, we show that the blue-velocity CO and H{\sc i} clouds are associated with the SNR, and the other clouds, at the red- and middle-velocities, are located in front of the SNR relative to the line of sight.

The middle-velocity ($V_{\rm LSR}$ $=$ 291--298 km s$^{-1}$) and red-velocity ($V_{\rm LSR}$ $=$ 298--306 km s$^{-1}$) H{\sc i} clouds show cavity-like structures in the position-velocity diagram, which have similar diameters to the SNR shell (Figure  \ref{fig4}). We also find H{\sc i}-dips toward the southeastern part of the N49 shell, where the radio continuum is bright [Figures  \ref{fig3}(b) and  \ref{fig3}(c)]. {We argue that since the H{\sc i} clouds in the middle- and red-velocities are located in front of the SNR, the H{\sc i}-dips in these clouds represent the absorption lines caused by the strong radio continuum radiation from the SNR as a background source} \citep[e.g.,][]{2013ApJ...768..179Y}. This means that both the middle- and red-velocity clouds are located on the nearside of the SNR {and not associated with the SNR}. For the blue-velocity component ($V_{\rm LSR}$ $=$ 280--291 km s$^{-1}$), there is no H{\sc i}-dip. This indicates that the blue-velocity component is not affected by the absorption effect, and is located on the far side of or within the SNR.

The X-ray studies also support this interpretation. According to X-ray spectroscopy, the absorbing column density for X-rays {$N_{\rm H}$} is $\sim$3.51 $\times$ 10$^{21}$ cm$^{-2}$ {estimated by the integrated spectrum of N49 with $Suzaku$ \citep{2015ApJ...808...77U}, while spatially resolved X-ray spectral model fits with $Chandra$ showed a range of $N_{\rm H} \sim$1--4 $\times$ 10$^{21}$ cm$^{-2}$ \citep{2003ApJ...586..210P, 2012ApJ...748..117P}. There may be a small-scale variation, or it may be model-dependent.} We can determine the total column density from both CO and H{\sc i} [i.e., $N_{\rm p}$(H$_2$+H{\sc i})] for the middle- and red-velocity components using the following equation:
\begin{equation}
  N_{\rm p}({\rm H}{\textsc i}) = 1.823 \times 10^{18} \cdot W({\rm H}{\textsc i})  ({\rm cm}^{-2}), 
\label{equ:4}
\end{equation}
\begin{equation}
  N_{\rm p}({\rm H_2}+{\rm H}{\textsc i}) = N_{\rm p}({\rm H}{\textsc i}) + 2 \times N_{\rm p}({\rm H_2}), 
\label{equ:5}
\end{equation}
where the $W$(H{\sc i}) is the integrated intensity of H{\sc i} in unit of  K km s$^{-1}$. Since middle- and red-velocity gases are absorbed in the direction of N49, we used the point ($\alpha_{\rm J2000}, \delta_{\rm J2000}$) $\sim$ ($5^{\mathrm h} 26^{\mathrm m} 00^{\mathrm s}, -66^{\mathrm d} 06^{\mathrm m} 15^{\mathrm s}$) where these gases are not absorbed to calculate $N_{\rm p}$(H{\sc i}) from equation (\ref{equ:4}). Thus, from equations (\ref{equ:3}), (\ref{equ:4}), and (\ref{equ:5}) we found $N_{\rm p}$(H$_2$ + H{\sc i}) to be $\sim$3 $\times$ 10$^{21}$ cm$^{-2}$, which is roughly consistent with the absorbing column density for X-rays.

We also find that the cloud positions relative to the line of sight are inconsistent with an expanding motion of the interstellar gas. It is {thought} that a strong stellar wind or accretion wind from the progenitor of a core-collapse or Type Ia SNR evacuates the ambient interstellar gas, which may thus be observed as an expanding gas motion \citep[e.g.,][]{2012ApJ...746...82F, 2017JHEAp..15....1S}. Blue- and red-shifted gases therefore represent the near and far sides of the cavity, respectively. In the case of N49, however, the blue-velocity cloud ($=$ blue-shifted cloud) is located at the far side of the SNR, which is the opposite of what is for expanding motions of the interstellar gas. This inconsistency can be understood if pre-existing gas motions dominate the expanding gas motion. The SNR N49 is located at the boundary between two supergiant shells, LMC 4 and 5, the molecular clouds of which were likely formed by the strong compression between them. The current expansion velocities of the two supergiant shells are $\sim$30--40 km s$^{-1}$ from the H{\sc i} data \citep{1999AJ....118.2797K, 2008ApJS..175..165B}, and hence, the pre-existing gas motions may be dominant in this region. 

The spatial distribution of the CO clumps provides further support for the idea that the blue-velocity component is associated with N49. We mapped the SNR with the $^{12}$CO($J$ = 1--0) emission line using Mopra and found that the molecular cloud at $V_{\rm LSR}$ = 281--291 km s$^{-1}$ lies along the southeastern shell of the SNR [see Figure \ref{fig3}(d)]. This is consistent with a previous study by \cite{{1997ApJ...480..607B}}, who used the SEST to observe the $^{12}$CO($J$ = 2--1) emission line. In addition to this, we found a good spatial correspondence between the CO and {soft} X-ray distributions, with a correlation coefficient of {$\sim$0.95, indicating that the surface of the CO clouds may be ionized by the shock interaction. Moreover, the {high-intensity} ratio of CO 3--2/1--0 in the blue-shifted CO cloud is a direct evidence for the shock-heating and/or shock-compression. Furthermore}, some of the CO clumps resolved by ALMA are rim-brightened in {both the soft and hard} X-rays, indicating that the origin of the X-rays may be physically connected with the CO clumps. {Since the age of SNR N49 $\sim 5000$ yr is relatively younger than that of the {middle-aged} SNRs (e.g., $\sim20000$ yr for W44, $\sim30000$ yr for IC~443), no clear evidence for the line broadening at the blue-velocity CO cloud is expected (see Figure \ref{spec}).} Finally, we conclude that both the blue-velocity CO and H{\sc i} clumps are likely to be associated with the SNR. 

We can also show that the spatial anti-correlation between the CO and hard X-ray peaks is not due to photometric absorption. According to \cite{1994hea2.book.....L}, the X-ray optical depth $\tau_{\rm x}$ is given by

\begin{equation}
  \tau_{\rm x} = 2 \times 10^{-22} N_{\rm p}({\rm H}_2+{\rm H}{\textsc i}) \cdot \varepsilon^{-8/3} ({\rm cm}^{-2}) ,  
\label{equ:6}
\end{equation}
where $\varepsilon$ is X-ray photon energy in unit of keV. For CO clump L, we found $N_{\rm p}$(H$_2$ + H{\sc i}) to be $\sim$4 $\times$ 10$^{21}$ cm$^{-1}$ using equations (\ref{equ:3})--(\ref{equ:5}). We therefore obtain the X-ray optical depths $\tau_{\rm x}$ are 5, 0.5, 0.1, and 0.005 at X-ray photon energies of 0.5, 1.2, 2, and 7 keV, respectively. In the hard X-ray band ($\varepsilon$ $=$ 2--7 keV), the X-ray optical depth is 0.005--0.1, and hence, the absorption does not significantly affect the hard-X-ray image. In contrast, the soft X-rays are significantly affected by absorption. For CO clump L, the soft X-rays are suppressed more than the hard X-rays [see Figure \ref{fig10}(c) at a distance of $\sim$4 pc]. The optical depth for soft X-rays is 0.5--5 in the energy band 0.5--1.2 keV, and hence, the soft X-rays are absorbed at the edge of CO clump L.

\subsection{{Physical conditions in the molecular clumps}}\label{subsec:phy}

We argue here that the 21 CO clumps associated with the SNR may not be in virial equilibrium. The total virial mass of the CO clumps is $\sim${5.3} $\times$ $10^3$ $M_{\sun}$, whereas the total CO-derived mass is $\sim${1.9} $\times$ $10^3$ $M_{\sun}$ (see Table \ref{tab1}), with the former being roughly three times higher than the latter. {It is possible that if a lower conversion factor between the $N$(H$_2$) and $W$(CO) was used, this difference between virial and CO mass would be even larger. On the other hand,} \cite{1997ApJ...480..607B} also mentioned this trend, using the SEST CO datasets on a 10 pc scale. They argued that shock waves might perturb the molecular clumps and that broad molecular emission lines (30--40 km s$^{-1}$) should be detected. We confirmed this trend on a pc scale; however, we cannot find clear evidence for the suggested line broadening in our ALMA CO datasets because of low sensitivity and low excitation line data. Further CO observations with high-excitation lines---e.g., the $^{12}$CO($J$ = 3--2) emission line---are needed to clarify the situation.

We note that the total mass of the CO clumps, $\sim${1.9} $\times$ $10^3$ $M_{\sun}$, is roughly consistent with that of a parent cloud that forms a single Galactic O-type star. According to \cite{2015ApJ...806....7T, 2017ApJ...835..142T}, the masses of such parent clouds are $\sim${1}--50 $\times$ $10^3$ $M_{\sun}$. To form a single O-type star, two molecular clouds must collide with each other at a velocity of $\sim$10 km s$^{-1}$. It is possible that the identified CO clumps are parts of the parent clouds that formed the massive progenitor of N49. If so, this would provide further support for a core-collapse origin for the SNR.

\newpage
\subsection{{Origins of the radio continuum and hard X-rays}}\label{subsec:hx}
We found that CO clumps J, K, and L are rim-brightened in both hard X-rays and the radio continuum (see Figures \ref{fig7} and \ref{fig9}). We here discuss the origins of the radio-continuum and hard X-ray enhancement around the CO clumps.

\subsubsection{{Magnetic field amplification via a shock-cloud interaction}}\label{subsec:sci}
The enhancement of synchrotron radiation around the CO clumps can be interpreted as a result of magnetic-field amplification via a shock-cloud interaction. According to \cite{2010ApJ...724...59S, 2013ApJ...778...59S, 2017JHEAp..15....1S}, the interstellar gas surrounding the Galactic SNRs RX J1713.7--3946 and RCW 86 is highly inhomogeneous and clumpy because the ambient gas was completely removed by strong stellar winds or by {optically-thick winds from the white dwarf when the mass accretion is effective (so-called ``accretion wind'')}, and dense gas clumps (such as the CO clumps) can survive such erosion. The density difference between the CO clumps and the inter-clump region is typically on the order of 10$^4$ \citep[e.g.,][]{2010ApJ...724...59S}. The inhomogeneous gas distribution generates {turbulent} motions around the CO clumps when the SNR shock waves {collide} with them. According to three-dimensional magnetohydrodynamic simulations, the turbulent motions enhance the magnetic-field strength up to $\sim$1 mG {\citep[][see also \citeauthor{2007ApJ...663L..41G} \citeyear{2007ApJ...663L..41G}, \citeauthor{2009ApJ...695..825I} \citeyear{2009ApJ...695..825I}, \citeauthor{2012ApJ...758..126S} \citeyear{2012ApJ...758..126S}, \citeauthor{2012ApJ...747...98G} \citeyear{2012ApJ...747...98G}]{2012ApJ...744...71I}}, and hence, the bright synchrotron radiation will be observed around the CO clumps. The typical spatial separation between the intensity peaks of the synchrotron X-ray/radio continuum and CO/H{\sc i} is 1.2 $\pm$ 0.6 pc for RX J1713.7--3946 \citep{2013ApJ...778...59S}, 1.8 $\pm$ 1.3 pc for the X-rays from RCW 86, and 1.4 $\pm$ 1.3 pc for the radio continuum from RCW 86 \citep{2017JHEAp..15....1S}.

For N49, the radio continuum represents the synchrotron radiation because the spectral index was measured to be $\sim$$-$0.59 $\pm$ 0.03 \citep{2017ApJS..230....2B}. Since we also found an interstellar cavity and dense molecular clumps, the interstellar environment of N49 is highly inhomogeneous. The spatial separation between the intensity peaks of the radio continuum and the CO in N49 is $\sim$1--3 pc (see Figures \ref{fig8} and \ref{fig9}), which is roughly consistent with the previous results for the Galactic SNRs RX J1713.7--3946 and RCW 86 \citep{2013ApJ...778...59S, 2017JHEAp..15....1S}. We therefore conclude that the radio continuum enhancement around the CO clumps in N49 is likely to be due to magnetic-field amplification via a shock-cloud interaction. Further detailed analysis is needed to clarify the physical properties of accelerated cosmic-ray electrons, e.g., by comparing the distribution of the spectral index with that of the interacting gas \citep[e.g.,][]{2015ApJ...799..175S}.

\subsubsection{{Thermal-conduction scenario for the recombining plasma}}\label{subsec:rcp}
As of early 2018, firm observational evidence for recombining plasma had been obtained for more than 10 Galactic SNRs (e.g., G359.1--0.5, G346.6--0.2, W28, W44, CTB37A, 3C391, MSH11--61A, IC443, W49B, and G166.0$+$4.3) and for the Magellanic SNR N49 \citep[see][and references therein]{2017PASJ...69...30M}. The origin of this recombining plasma is one of the major issues in modern X-ray astrophysics. {Here we propose that the interaction of hot gas with the cold and dense molecular clumps results in the creation of recombining plasma in N49.}

{For N49, \cite{2015ApJ...808...77U} revealed significant residuals of the spectral modeling at 2.0 keV and 2.6 keV from the collisional ionization equilibrium (CIE) plasma of 0.6 keV, which represent to the centroid energies of the Si Ly${\alpha}$ and S Ly${\alpha}$ emission lines, respectively. Furthermore, a hump-like feature is seen in the residual of the spectral modeling at $\sim2.7$ keV, indicating a strong evidence for the radiative recombination continua (RRC). The authors concluded that the residuals could be reproduced the recombining plasma. Therefore, the hard X-ray image in the energy band of 2.0--7.0 keV can be traced a spatial distribution of Si Ly${\alpha}$ emission and RRC.}

{We revealed that} the hard X-rays are significantly enhanced around CO clumps J, K, and L (Figure \ref{fig8}). CO clump {L} is fully surrounded by hard X-rays, and the CO clumps {J and K} are bright in hard X-rays only on one side. The spatial separation between the intensity peaks of the CO and X-rays is $\sim$0.7 pc for CO clump {L} and $\sim$2 pc for CO clumps {J and K} (see Figure \ref{fig9}). We infer that the CO clump {L} has only recently started to interact with the shock wave. It is therefore likely that the hard-X-ray enhancement is related to the interaction between the SNR shock and the CO clumps.

The thermal-conduction scenario may explain the origin of the hard X-rays in N49. According to \cite{2002ApJ...572..897K}, when an expanding ionized plasma interacts with a cold molecular cloud, the electron temperature falls, and a recombining plasma is thus formed. In this scenario, one expects to observe hard X-rays being dominated by recombining plasma around shock-interacting molecular clouds. The spatial anti-correlation between the CO clumps and the hard X-rays in N49 supports the possibility that the recombining plasma was formed by the cold clouds. To confirm this scenario, {we should reveal spatial distributions of the electron temperature and the ionization timescale using $Chandra$ datasets, and to compare them with our ALMA CO distribution \citep[e.g.,][]{2017ApJ...851...73M,2018PASJ..tmp...29O}.}

We also note that numerical calculations that describe the physical properties of the neutral gas, thermal plasma, and relativistic particles in a shock-interaction region are required to confirm not only the thermal-conduction scenario but also the shock-cloud interaction. During a shock-cloud interaction, the electron temperature decreases, and the surface temperatures of molecular clumps increase. However, no study has yet attempted to clarify the physical properties and their time evolution for an inhomogeneous interstellar environment \citep[e.g.,][]{2012ApJ...744...71I, 2017ApJ...846...77S}. In addition, the time evolution of the thermal and non-thermal components is still unknown. In the free-expansion phase, cosmic-ray electrons are significantly accelerated by the shock waves with velocities of $\sim$1000 km s$^{-1}$, and non-thermal X-rays are brighter than the thermal X-rays. By several thousand years after the explosion, however, the non-thermal X-ray flux has decreased greatly because of the synchrotron cooling and shock deceleration, and hence, the thermal X-rays are dominant, as in the SNR N49. To quantify these qualitative descriptions, we need to determine the balance between particle acceleration and thermalization of the neutral gas and determine the particle energy-distribution functions.

\section{Summary} \label{sec:sum}
We have made new observations of molecular and atomic gas toward the Magellanic SNR N49 with Mopra, {ASTE,} ALMA, and ATCA using the $^{12}$CO($J$ = 1--0{, 3--2}) and H{\sc i} lines. The main results and conclusions of our study can be summarized as follows;

\begin{enumerate}
 \item We have categorized the H{\sc i}/CO clouds toward the SNR N49 into three velocity components using the spectral profiles obtained {using} Mopra and ATCA {with angular resolutions of $\sim20''$--$45''$.} The middle-velocity ($V_{\rm LSR}$ = 291--298 km s$^{-1}$) and red-velocity ($V_{\rm LSR}$ = 298--306 km s$^{-1}$) H{\sc i} clouds show dip-like structures in the direction of the southeastern parts of the SNR shell and have cavity-like structures in the position-velocity diagram, suggesting that the {H{\sc i}-dips in these clouds represent the absorption lines caused by the strong radio continuum radiation from the SNR as a background source.} We therefore conclude that both the middle- and red-velocity clouds are located in front of the SNR relative to the line of sight{, indicating that the gas motion is opposite than what is expected for the expanding motion of the interstellar gas.} The total column density of gas for these two clouds is also consistent with the absorbing column density derived from the X-ray spectroscopy \citep{2003ApJ...586..210P, 2012ApJ...748..117P, 2015ApJ...808...77U}.  The blue-velocity CO and H{\sc i} clouds lie along the southeastern shell of the SNR, where both the {X-rays} and radio continuum are bright. {The CO $J$ = 3--2 / 1--0 ratio indicates higher values of the blue-velocity CO cloud toward the SNR shell. Thus, strong interaction was suggested.} We conclude that the blue-{velocity} CO/H{\sc i} clouds are likely to be associated with N49.

 \item Using ALMA data, we have resolved the blue-velocity CO cloud spatially into 21 CO clumps with radii of 1--2 pc {with an angular resolution of $\sim3''$.} The total virial mass is $\sim${5.3} $\times$ $10^3$ $M_{\sun}$, although the CO-derived mass is only $\sim${1.9} $\times$ $10^3$ $M_{\sun}$, indicating that the SNR shock wave may be perturbing the molecular clumps. However, we observe no {significant} broadening of the molecular lines {due to the young age of the SNR}. We note that the total mass of the CO clumps is roughly consistent with that of a parent cloud that forms a single Galactic O-type star \citep[$\sim$1--50 $\times$ $10^3$ $M_{\sun}$;][]{2015ApJ...806....7T, 2017ApJ...835..142T}, providing further support for a core-collapse origin for N49.

 \item  We have found that some of the CO clumps (J, K, and L) that are resolved by ALMA are rim-brightened in both hard X-rays and the radio continuum on a 0.7--2 pc scale. This provides further evidence for dynamical interactions between the CO clumps and the shocks. We also note that the spatial anti-correlation between the CO and hard X-ray peaks is not due to photometric absorption, whereas the soft X-rays are likely to be absorbed by the CO clumps because of the large X-ray optical depths, $\sim$0.5--5 \citep[c.f.,][]{1994hea2.book.....L} for an interstellar gas column density of $\sim$4 $\times$ 10$^{21}$ cm$^{-2}$.

 \item We have argued that the enhancement of the synchrotron radio continuum around the CO clumps can be interpreted as a result of magnetic-field amplification via a shock-cloud interaction. The spatial separation between the intensity peaks of the radio synchrotron and CO in N49 is $\sim$1--3 pc, which is roughly consistent with previous results for the Galactic SNRs RX J1713.7$-$3946 and RCW~86 \citep{2013ApJ...778...59S, 2017JHEAp..15....1S}. Further detailed analysis is needed to clarify the physical properties of accelerated cosmic-ray electrons. i.e., by comparison between the spatial distribution of the spectral index with that of the interstellar gas \citep[e.g.,][]{2015ApJ...799..175S}.

 \item We have found that recombining plasma that dominates in hard X-rays is clearly enhanced around CO clumps J, K, and L on a 0.7--2.0 pc scale. A possible scenario is that the recombining plasma was formed via {interaction} between the SNR shock wave and the cold/dense molecular clumps \citep[e.g.,][]{2002ApJ...572..897K}. To confirm this scenario, {spatial comparisons among the interstellar gas, electron temperature, and  the ionization timescale are also needed \citep[e.g.,][]{2017ApJ...851...73M,2018PASJ..tmp...29O}.} 

\end{enumerate}

\acknowledgments

{\footnotesize{{We are grateful to Aya Bamba, Takaaki Tanaka, Hiroyuki Uchida, and Hideaki Matsumura for thoughtful comments and their contribution on the X-ray properties.} This paper makes use of the following ALMA data:ADS/JAO. ALMA\#2015.1.01195.S. ALMA is a partnership of ESO (representing its member states), NSF (USA) and NINS (Japan), together with NRC (Canada) and NSC and ASIAA (Taiwan) and KASI (Republic of Korea), in cooperation with the Republic of Chile. The Joint ALMA Observatory is operated by ESO, AUI/NRAO and NAOJ. The ASTE telescope is operated by National Astronomical Observatory of Japan (NAOJ). The scientific results reported in this article are based on data obtained from the $Chandra$ Data Archive (Obs IDs 10123, 10806, 10807, and 10808, PI: S. Park). This research has made use of software provided by the $Chandra$ X-ray Center (CXC) in the application packages CIAO (v 4.9). The Australia Telescope Compact Array, Parkes radio telescope, and Mopra radio telescope are part of the ATNF which is funded by the Australian Government for operation as a National Facility managed by CSIRO. This study was financially supported by Grants-in-Aid for Scientific Research (KAKENHI) of the Japanese Society for the Promotion of Science (JSPS, grant Nos. 15H05694 and 16K17664). {K. Tokuda was supported by NAOJ ALMA Scientific Research Grant Number 2016-03B.} {HS} was supported by ``Building of Consortia for the Development of Human Resources in Science and Technology'' of Ministry of Education, Culture, Sports, Science and Technology (MEXT, grant No. 01-M1-0305). {ML acknowledges support from the Ministry of Education, Science and Technological Development of the Republic of Serbia through project No. 176001. }{We are grateful to the anonymous referee for useful comments which helped the authors to improve the paper.}}

{\bf{\software{CASA \citep [v 4.7.2;][]{2007ASPC..376..127M}, MIRIAD \citep{1995ASPC...77..433S}, CIAO \citep[v 4.9;][]{2006SPIE.6270E..1VF}}}}

\clearpage

\begin{deluxetable*}{ccccccrrr}
\tablewidth{\linewidth}
\tabletypesize{\scriptsize}
\tablenum{1}
\tablecaption{Physical Properties of $^{12}$CO($J$ = 1--0) Clumps}
\label{tab1}
\tablehead{\\
\multicolumn{1}{c}{Clump name} & $\alpha$(2000) & $\delta$(2000) & $T_\mathrm{R}^{\ast}$ & $V_{\mathrm{peak}}$ &  $\bigtriangleup V$ & \multicolumn{1}{c}{Size}  & \multicolumn{1}{c}{$M_{\mathrm{vir}}$} & \multicolumn{1}{c}{$M_{\mathrm{CO}}$} \\
 & (h m s) & ($^{\circ}$ $'$ $''$) & (K) & (km $\mathrm{s^{-1}}$) & (km $\mathrm{s^{-1}}$)  & \multicolumn{1}{c}{(pc)} & \multicolumn{1}{c}{($M_\sun $)} & \multicolumn{1}{c}{($M_\sun $)} \\
\multicolumn{1}{c}{(1)} & (2) & (3) & (4) & (5) & (6) & \multicolumn{1}{c}{(7)} & \multicolumn{1}{c}{(8)} & \multicolumn{1}{c}{(9)}
}
\startdata
{A} & 5$:$26$:$5.4 & $-$66$:$5$:$37 & 5.3 & 283.6 & 1.6  & 1.4  &  $400$  &  $120$  \\
{B} & 5$:$26$:$5.5 & $-$66$:$5$:$50 & 5.1 & 284.2 & 1.1 &  0.9 &  $110$ &  $30$ \\
{C} & 5$:$26$:$5.0 & $-$66$:$5$:$4 6  & 2.5  & 284.3  & 1.4  & 1.5  &  $320$  &  $60$  \\
{D} & 5$:$26$:$4.9 & $-$66$:$5$:$34 &  5.9 & 284.5 & 1.2 & 1.4 &  $210$ &  $100$ \\
{E} & 5$:$26$:$8.5 & $-$66$:$5$:$30 & 2.2 & 284.8 & 1.4 & 1.6 &  $360$ &  $70$ \\
{F} & 5$:$26$:$8.3 & $-$66$:$5$:$36 & 2.0 & 285.0 & 0.8 &  0.9 &  $50$ & $10$ \\
{G} & 5$:$26$:$8.6 & $-$66$:$5$:$23 & 4.7 & 285.3 & 1.0 & 1.6 &  $130$ &  $50$ \\
{H} & 5$:$26$:$7.2 & $-$66$:$5$:$20 & 3.5 & 285.4 & 0.9 & 1.2 &  $100$ &  $30$ \\
{I} & 5$:$26$:$8.0 & $-$66$:$5$:$17 &  4.4 & 285.6 & 0.9 & 1.5 &  $110$ &  $70$ \\
{J} & 5$:$26$:$1.6 & $-$66$:$5$:$29 & 4.3 & 285.8 & 1.7 & 1.2 &  $350$ & $70$ \\
{K} & 5$:$26$:$5.5 & $-$66$:$5$:$21 & 9.2 & 286.0 & 2.0 & 2.0 & {$810$} & {$370$} \\
{L} & 5$:$26$:$4.7 & $-$66$:$5$:$02 & 5.8 & 286.0 & 1.6 & 1.5 &  $400$ &  $140$ \\
{M} & 5$:$26$:$3.7 & $-$66$:$4$:$47 & 2.1 & 286.1 & 0.7 & 1.1 &  $60$ & $10$ \\
{N} & 5$:$26$:$6.7 & $-$66$:$5$:$15 &  6.7 & 286.2 & 1.1 & 1.6 &  $200$ &  {$150$} \\
{O} & 5$:$26$:$3.7 & $-$66$:$5$:$23 & 11.5 & 286.2 & 1.1 & 1.4 &  $160$ &  $160$ \\
{P} & 5$:$26$:$7.5 & $-$66$:$5$:$12 & 6.3 & 286.3 &  1.3 & 1.1 &  $210$ &  $70$ \\
{Q} & 5$:$26$:$1.7 & $-$66$:$5$:$34 & 5.1 & 286.3 &  0.9 & 1.1 &  $100$ &  $40$ \\
{R} & 5$:$26$:$6.3 & $-$66$:$5$:$11 & 7.5 & 286.4 & 1.3 & 1.4 &  $250$ &  $150$ \\
{S} & 5$:$26$:$4.8 & $-$66$:$5$:$15 & 4.2 & 286.5 & 2.3 & 1.3 &  $720$ &  $110$ \\
{T} & 5$:$26$:$1.0 & $-$66$:$5$:$27 & 4.9 & 286.6 &  0.9 & 0.9 &  $80$ & $30$ \\
{U} & 5$:$26$:$1.9 & $-$66$:$4$:$53 & 1.4 & 287.4 & 1.4 & 1.0 &  $200$ &  $10$ \\
\enddata
\tablecomments{Col. (1): Clump name. Cols. (2--3): Position of the maximum CO intensity for each velocity component. Cols. (4--6): Physical properties of the $^{12}$CO($J$ = 1--0) emission obtained at each position. Col. (4): Peak radiation temperature, $T_\mathrm{R}^{\ast} $. Col. (5): $V_{\mathrm{peak}}$ derived from a single Gaussian fitting. Col. (6): Full-width at half-maximum (FWHM) line width, $\bigtriangleup V$. Col. (7): Cloud size defined as ($A$/$\pi$)$^{0.5}$ $\times 2 $, where $A$ is the total cloud surface area surrounded by the half of peak radiation temperature contour (see the text). Col. (8): Mass of the cloud derived using the virial theorem. Col. (9): Mass of the cloud derived by using the relationship between the molecular hydrogen column density $N$(H$_{2}$) and the $^{12}$CO($J$ = 1--0) intensity $W(^{12}$CO)$/N$(H$_2$)=7.0${\times}$10$^{20}$ [$W$($^{12}$CO)$/$(K km s$^{-1}$)]cm$^{-2}$) \citep{2008ApJS..178...56F}.}
\end{deluxetable*}

\begin{figure*}
\begin{center}
\includegraphics[width=90mm,clip]{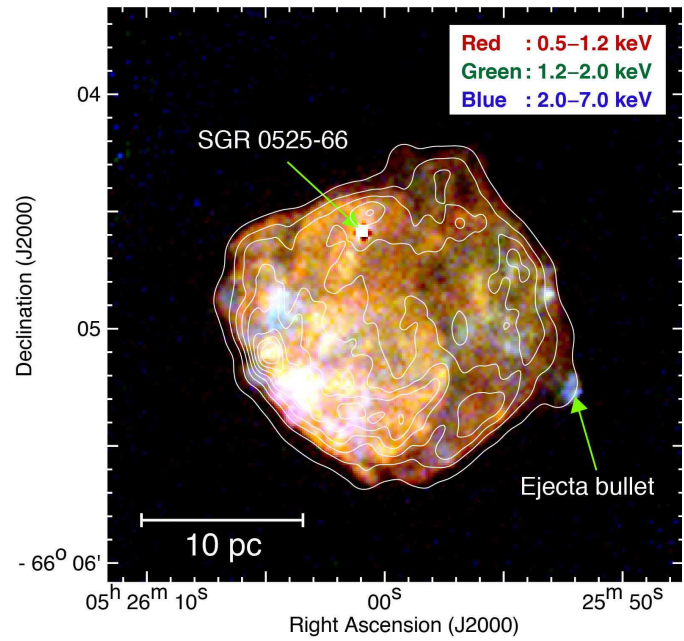}
\caption{Three-color $Chandra$ ACIS image of N49 {\citep{2012ApJ...748..117P}}. Red, green, and blue correspond to the energy bands of  0.5--1.2 keV (soft), 1.2--2.0 keV (medium), and 2.0--7.0 keV (hard), respectively. White contours indicate the radio continuum (1.42 GHz) delineated every 2.0 $\times$ 10$^{-3}$ Jy beam$^{-1}$ from 0.8 $\times$ 10$^{-4}$ Jy beam$^{-1}$.}
\label{fig1}
\end{center}
\end{figure*}

\begin{figure*}
\begin{center}
\includegraphics[width=\linewidth,clip]{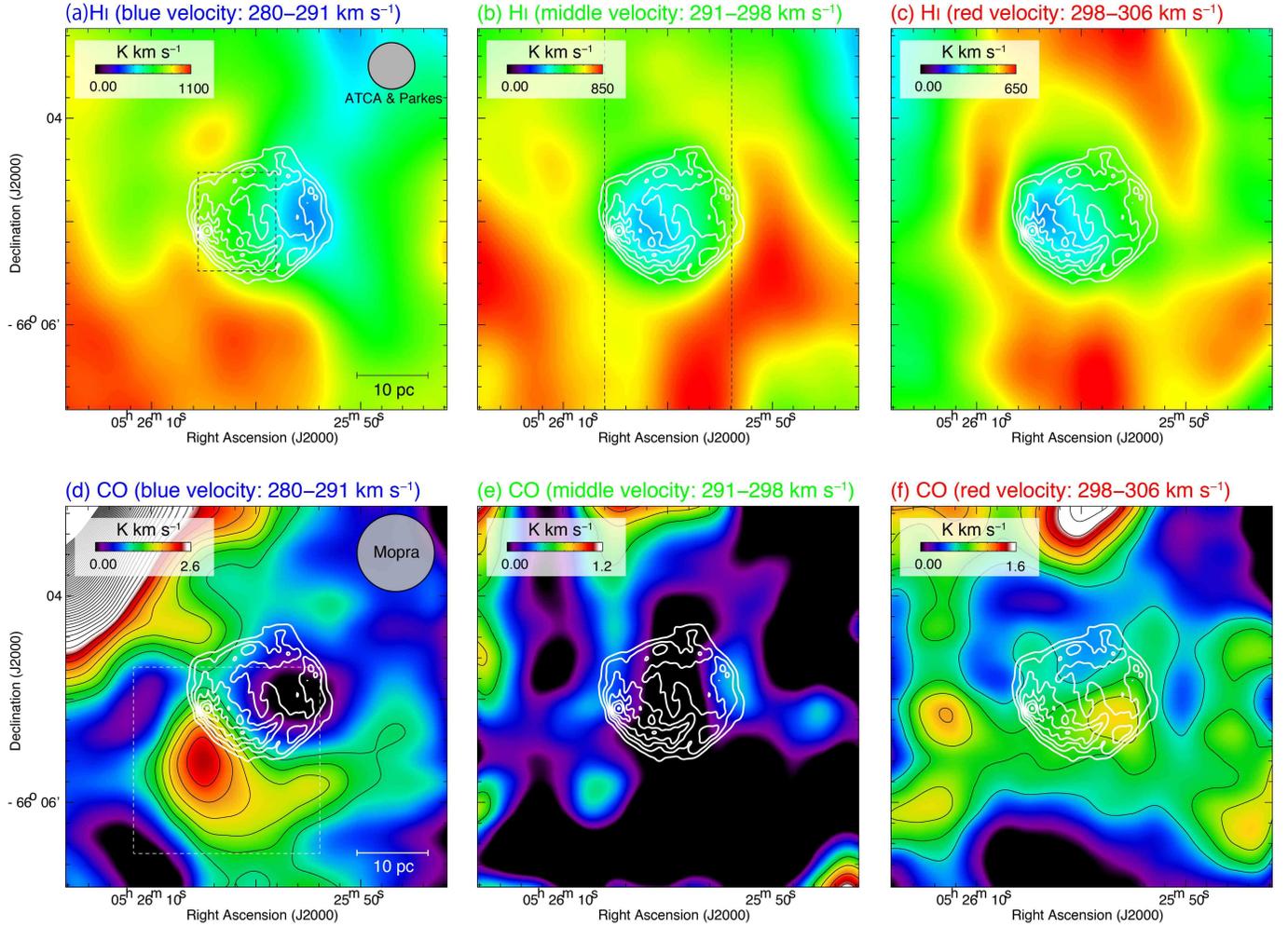}
\caption{{Integrated intensity maps of (a--c) ATCA $\&$ Parkes H{\sc i} and (d--f) Mopra $^{12}$CO$(J$ = 1--0). The integration velocity range is $V_{\rm LSR}$ $=$ 280--291 km s$^{-1}$ (blue-velocity) for (a) and (d); $V_{\rm LSR}$ $=$ 291--298 km s$^{-1}$ (middle-velocity) for (b) and (e); $V_{\rm LSR}$ $=$ 298--306 km s$^{-1}$ (red-velocity) for (c) and (f).} White contours indicate the radio continuum (1.42 GHz) delineated every 2.5 $\times$ 10$^{-3}$ Jy beam$^{-1}$ from 2.0 $\times$ 10$^{-3}$ Jy beam$^{-1}$. Black contours {indicate CO integrated intensities} delineated every 0.21 K km s$^{-1}$ from 1.1 K km s$^{-1}$ for (d); every 0.19 K km s$^{-1}$ from 0.56 K km s$^{-1}$ for (e) and (f).}
\label{fig3}
\end{center}
\end{figure*}%

\begin{figure*}
\begin{center}
\includegraphics[width=90mm,clip]{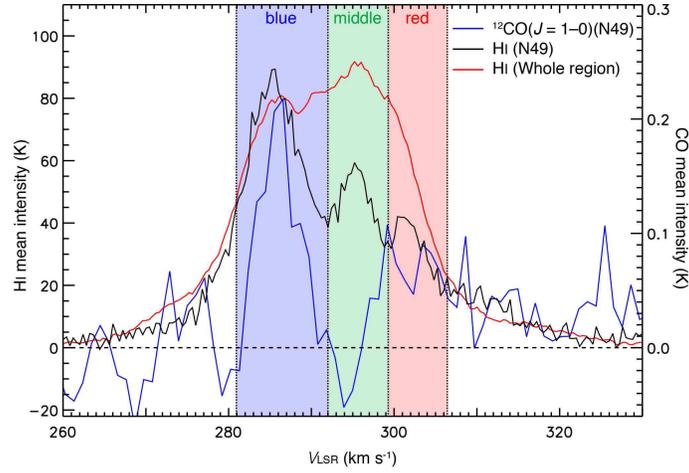}
\caption{H{\sc i} and CO spectra in N49. Black line is ATCA $\&$ Parkes H{\sc i} spectrum averaged over the rectangle region as shown in Figure \ref{fig3}(a). Blue line is Mopra $^{12}$CO$(J$ = 1--0) spectrum averaged over the rectangle region as shown in Figure \ref{fig3}(d). Red line is ATCA $\&$ Parkes H{\sc i} spectrum averaged over the whole region of Figure \ref{fig3}. {The blue, green, and red-shaded areas represent the blue-velocity ($V_{\rm LSR}$ $=$ 280--291 km s$^{-1}$), the middle-velocity ($V_{\rm LSR}$ $=$ 291--298 km s$^{-1}$), and the red-velocity ($V_{\rm LSR}$ $=$ 298--306 km s$^{-1}$), respectively.}}
\label{fig2}
\end{center}
\end{figure*}%

\begin{figure*}
\begin{center}
\includegraphics[width=90mm,clip]{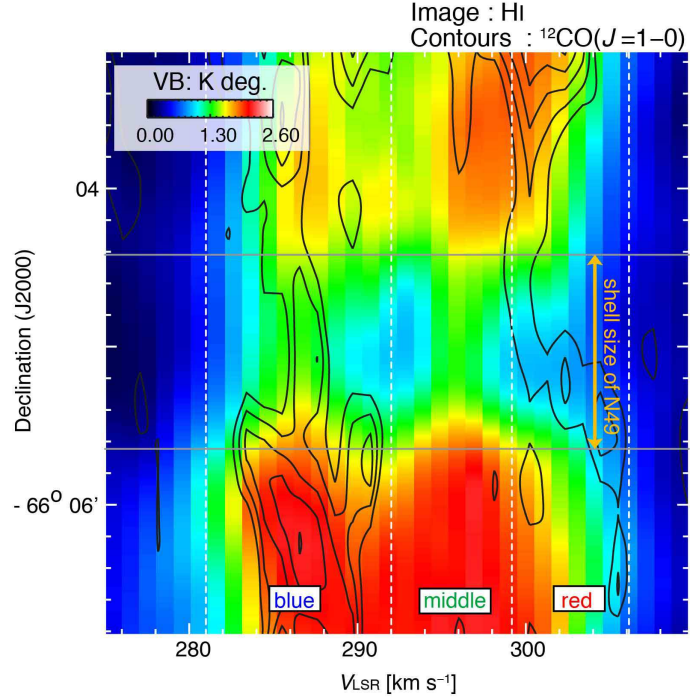}
\caption{Position-velocity diagram of the ATCA $\&$ Parkes H{\sc i} image and the Mopra $^{12}$CO$(J$ = 1--0) contours. The integration range is from 81$\fdg$47 to 81$\fdg$53 in the Right Ascension. Orange arrow and interval of two {gray} lines indicate the diameter with respect to N49 in terms of the Declination range. Black contours indicate the Mopra $^{12}$CO$(J$ = 1--0) delineated every 9.0 $\times$ 10$^{-4}$ K km s$^{-1}$ from 1.8 $\times$ 10$^{-3}$ K km s$^{-1}$.}
\label{fig4}
\end{center}
\end{figure*}%

\begin{figure*}
\begin{center}
\includegraphics[width=\linewidth,clip]{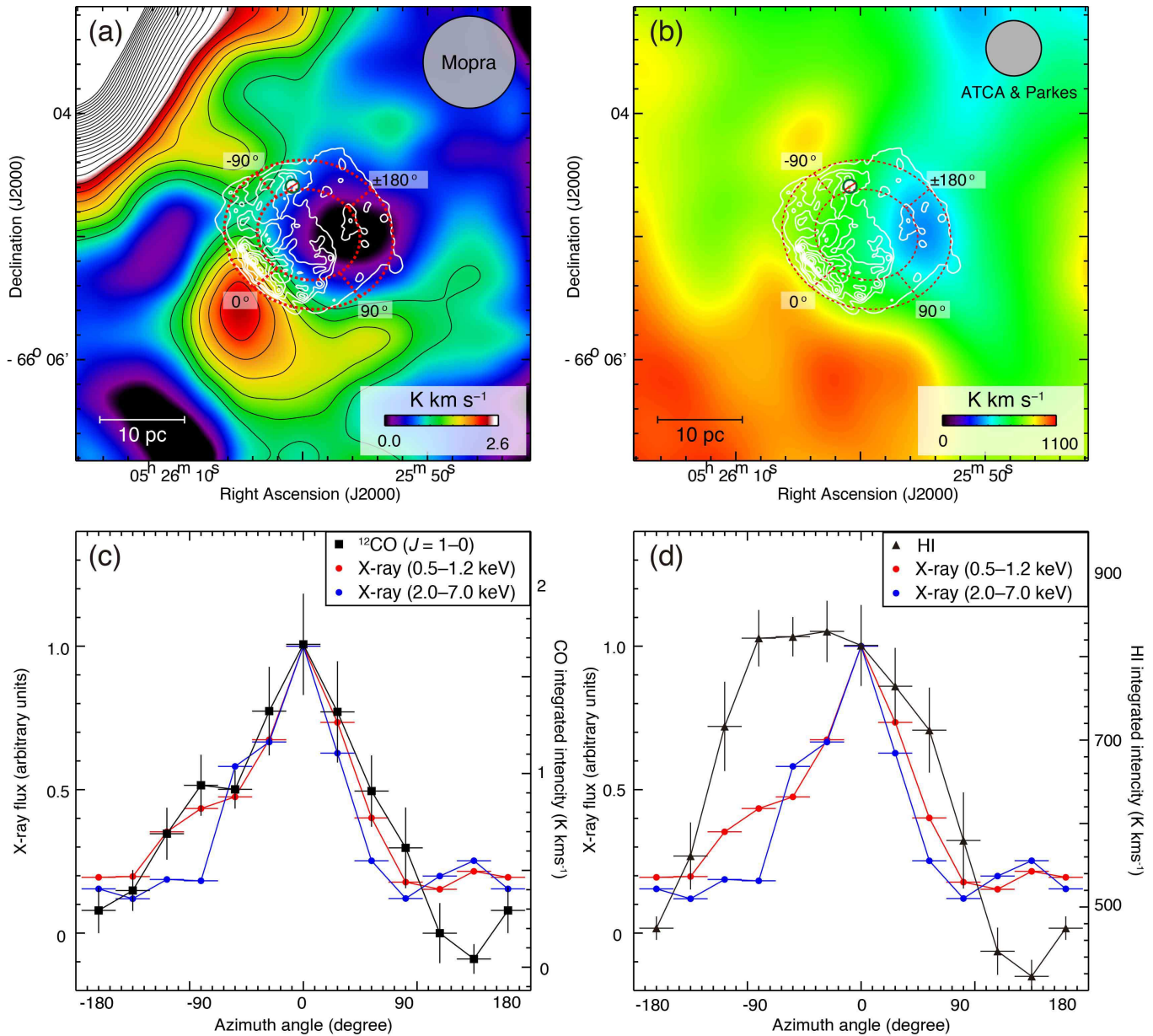}
\caption{
{(a--b)} Spatial distribution of (a) Mopra $^{12}$CO$(J$ = 1--0) and (b) ATCA $\&$ Parkes H{\sc i}. These velocity ranges are the same as in Figures \ref{fig3}(a) and \ref{fig3}(d). These contours are {broad-band $Chandra$ X-ray} ($\varepsilon = 0.5$--7.0 keV) plotted at every {5 $\times$ 10$^{-7}$} photons pixel$^{-1}$ s$^{-1}$ from {1 $\times$ 10$^{-7}$} photons pixel$^{-1}$ s$^{-1}$. {(c--d)} Azimuthal distributions of $Chandra$ {broad band} and soft X-rays compare with (c) the Mopra $^{12}$CO$(J$ = 1--0) and (d) the ATCA $\&$ Parkes H{\sc i} between the elliptical rings shown in (a) and (b). Semimajor and semiminor radius of outer ring are 10 pc and 9 pc, respectively. The direction where the X-ray flux becomes maximum is defined as the origin of the azimuth angle, which is measured counterclockwise. {The X-rays from SGR 0525--66 annotated by the green circles were excluded from this analysis.}}
\label{fig5}
\end{center}
\end{figure*}%

\begin{figure*}
\begin{center}
\includegraphics[width=90mm,clip]{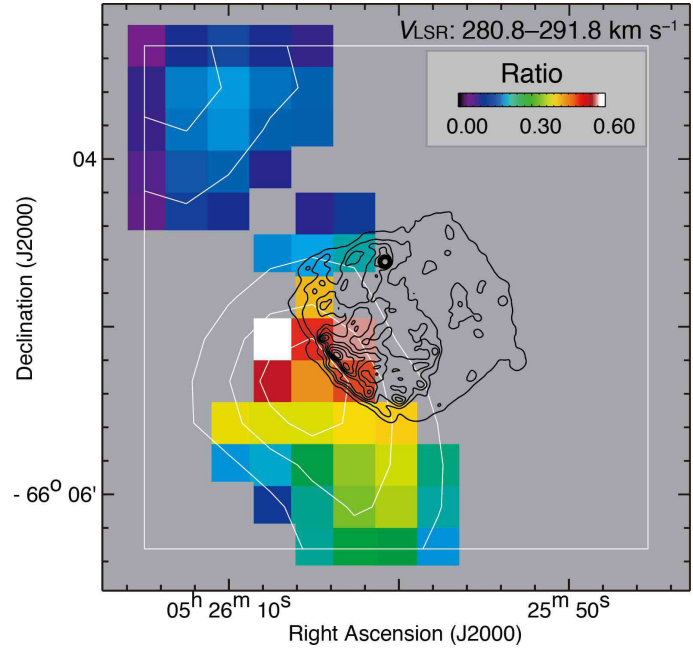}
\caption{{Distribution of the $^{12}$CO$(J$ = 3--2)/($J$ = 1--0) intensity ratio using the ASTE and Mopra data sets at the blue velocity. The contours indicate {broad-band $Chandra$ X-rays} and the contour levels are the same as in Figures \ref{fig5}a and \ref{fig5}b. White contours indicate the intensity of ASTE $^{12}$CO$(J$ = 3--2) delineated every 0.27 K km s$^{-1}$ ($\sim3\sigma$).}}
\label{ratio}
\end{center}
\end{figure*}%

\begin{figure*}
\begin{center}
\includegraphics[width=\linewidth,clip]{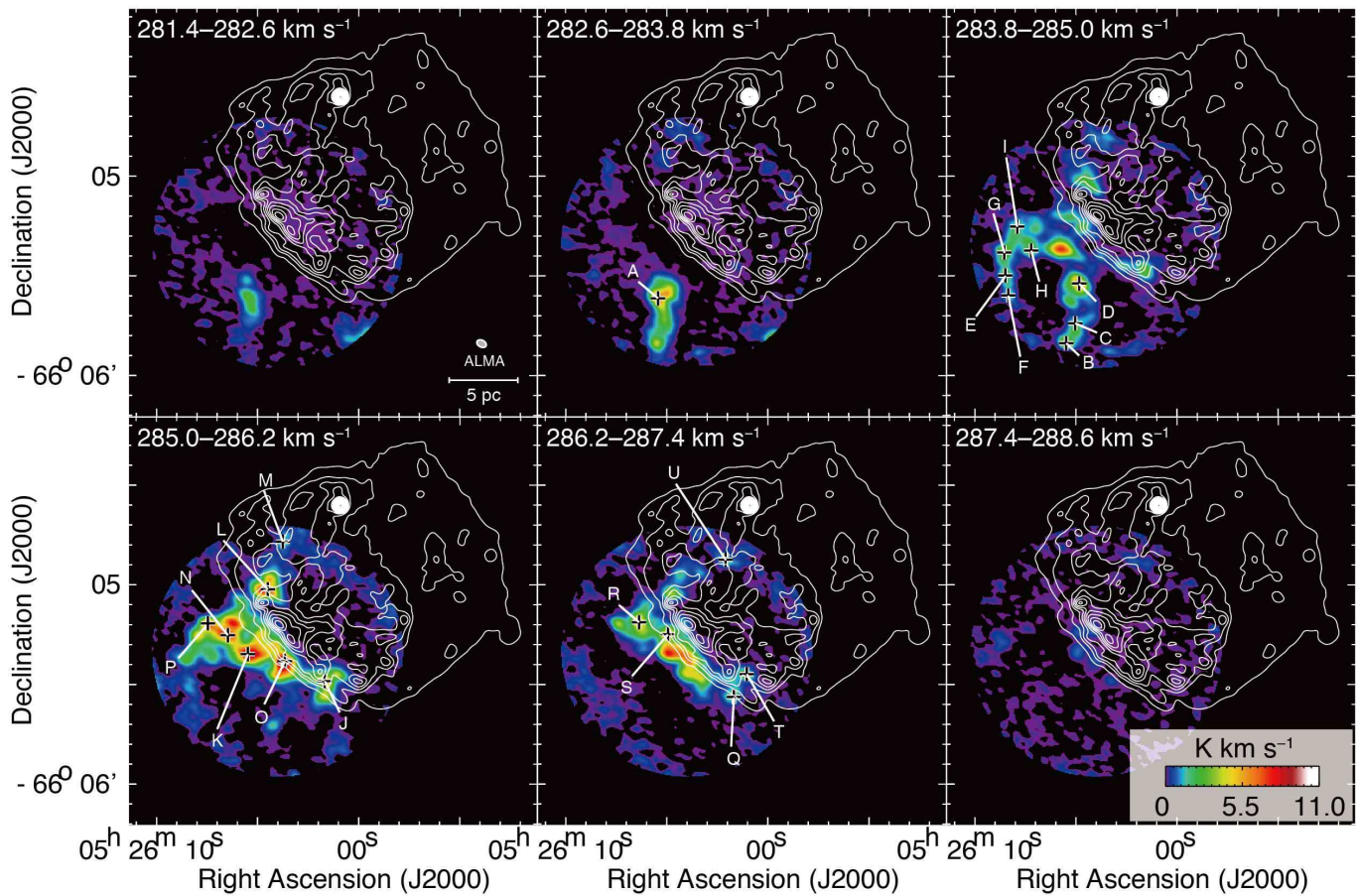}
\caption{ALMA $^{12}$CO$(J$ = 1--0) channel maps overlaid on the {hard X-rays $Chandra$} contours toward N49. Each panel shows the CO intensity map integrated over the velocity range from 281.4 to 288.6 km s$^{-1}$ every 1.2 km s$^{-1}$. {White} contours of X-rays are the same as in Figure \ref{fig5}. The CO clumps (A--U) are detailed in Table \ref{tab1}.}
\label{fig6}
\end{center}
\end{figure*}%

\begin{figure}
\begin{center}
\includegraphics[width=70mm,clip]{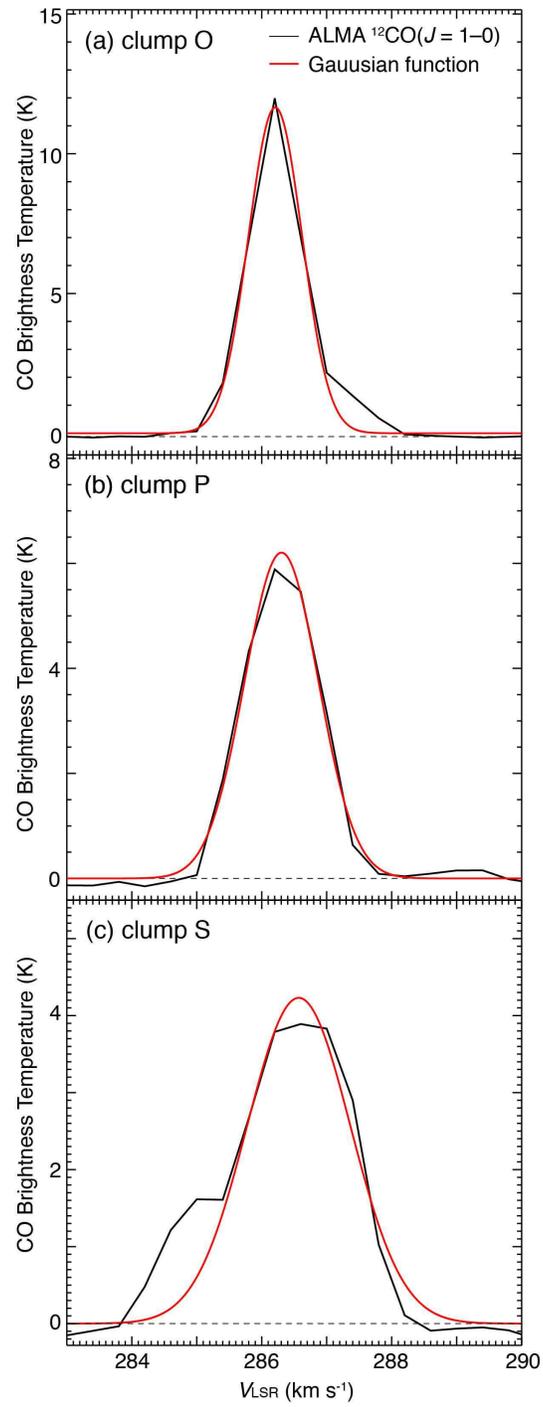}
\caption{{CO spectra in (a) clump O, (b) clump P, and (c) clump S obtained using ALMA. The red lines indicate fitting results using a Gaussian function.}}
\label{spec}
\end{center}
\end{figure}%

\begin{figure*}
\begin{center}
\includegraphics[width=\linewidth,clip]{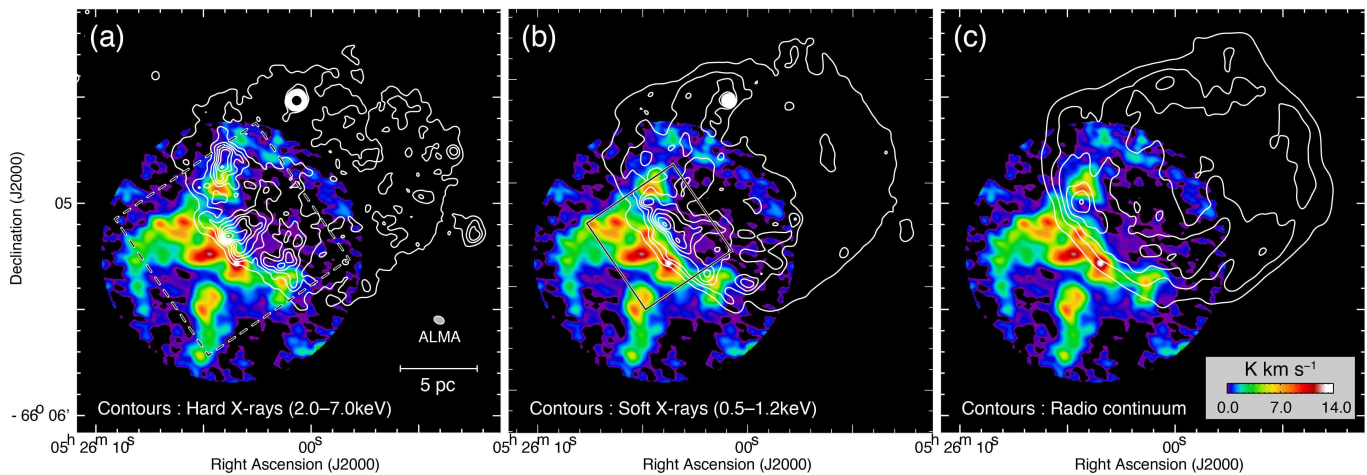}
\caption{ALMA $^{12}$CO$(J$ = 1--0) integrated intensity maps. The integrated velocity range corresponds to the blue component. White contours indicate the hard X-rays (the same as in Figure \ref{fig6}) in (a), and the soft X-rays plotted at every 4 $\times$ 10$^{-7}$ photons pixel$^{-1}$ s$^{-1}$ from 1 $\times$ 10$^{-7}$ photons pixel$^{-1}$ s$^{-1}$ in (b), and the radio continuum delineated every 3 mJy beam$^{-1}$ from 2 mJy beam$^{-1}$ in (c).}
\label{fig7}
\end{center}
\end{figure*}%

\begin{figure*}
\begin{center}
\includegraphics[width=\linewidth,clip]{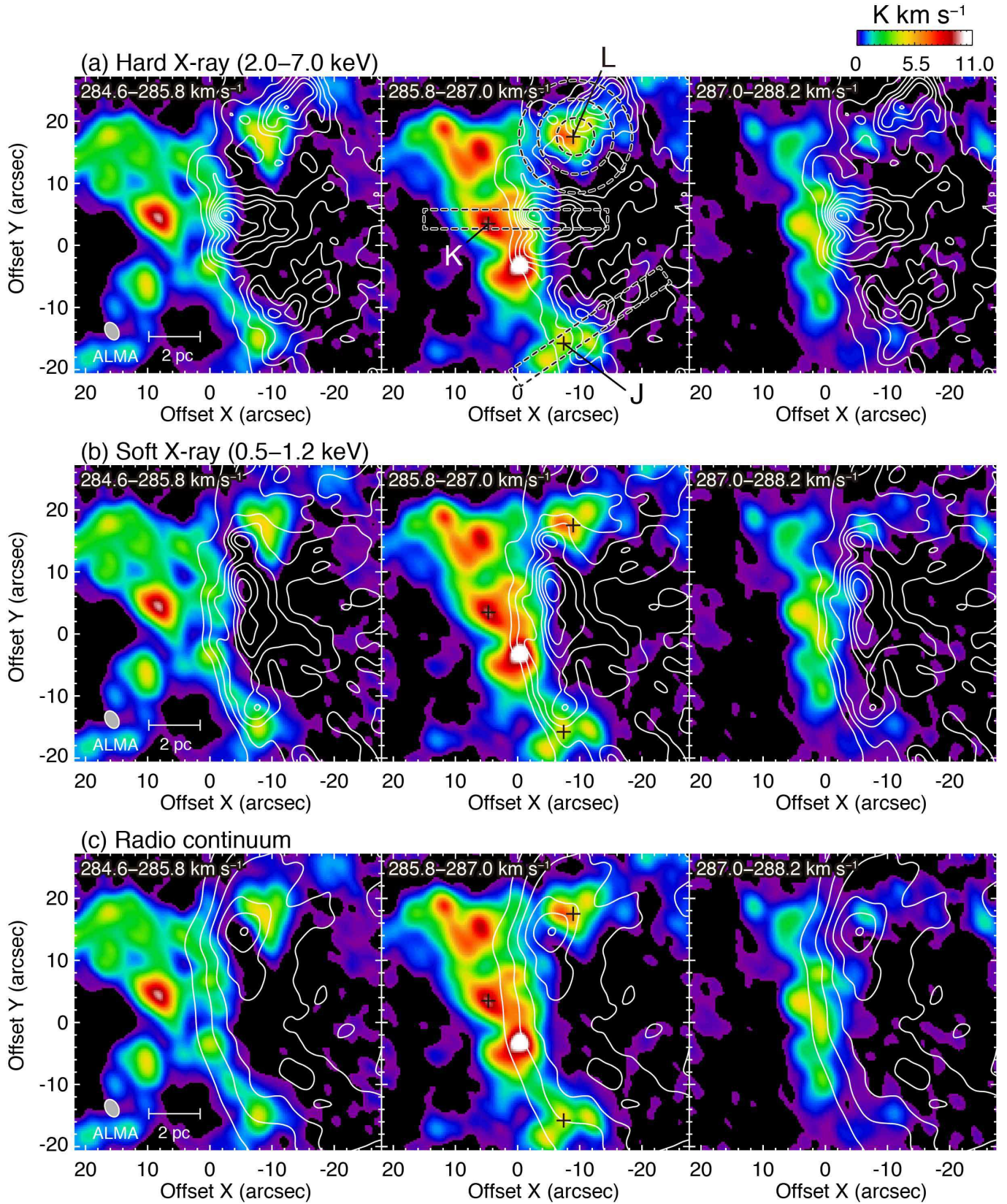}
\caption{ALMA $^{12}$CO$(J$ = 1--0) channel maps toward dashed regions in Figure \ref{fig7}(a). Each panel shows the CO intensity map integrated over the velocity range from 284.6 to 288.2 km s$^{-1}$ every 1.2 km s$^{-1}$. {White contours} in (a), (b), and (c) are the same as in Figures \ref{fig7}(a), \ref{fig7}(b), and \ref{fig7}(c), respectively. {The crosses indicate the positions of Clumps J, K, and L.}	}
\label{fig8}
\end{center}
\end{figure*}%

\begin{figure*}
\begin{center}
\includegraphics[width=\linewidth,clip]{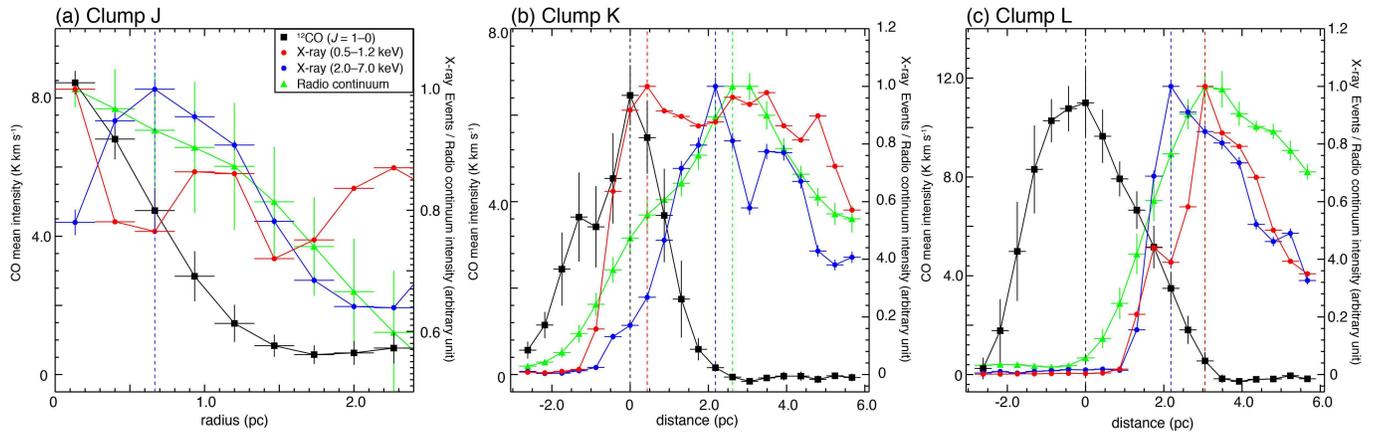}
\caption{Radial Profiles of the $^{12}$CO$(J$ = 1--0) line emission (black), soft X-rays (red), hard X-rays (blue), and radio continuum (green) for the circular or rectangular regions as shown in Figure \ref{fig8}(a). Black, red, blue, and green dashed lines indicate the intensity peaks of $^{12}$CO$(J$ = 1--0) emission, soft X-rays, hard X-rays, and radio continuum, respectively.}
\label{fig9}
\end{center}
\end{figure*}%

\begin{figure*}
\begin{center}
\includegraphics[width=\linewidth,clip]{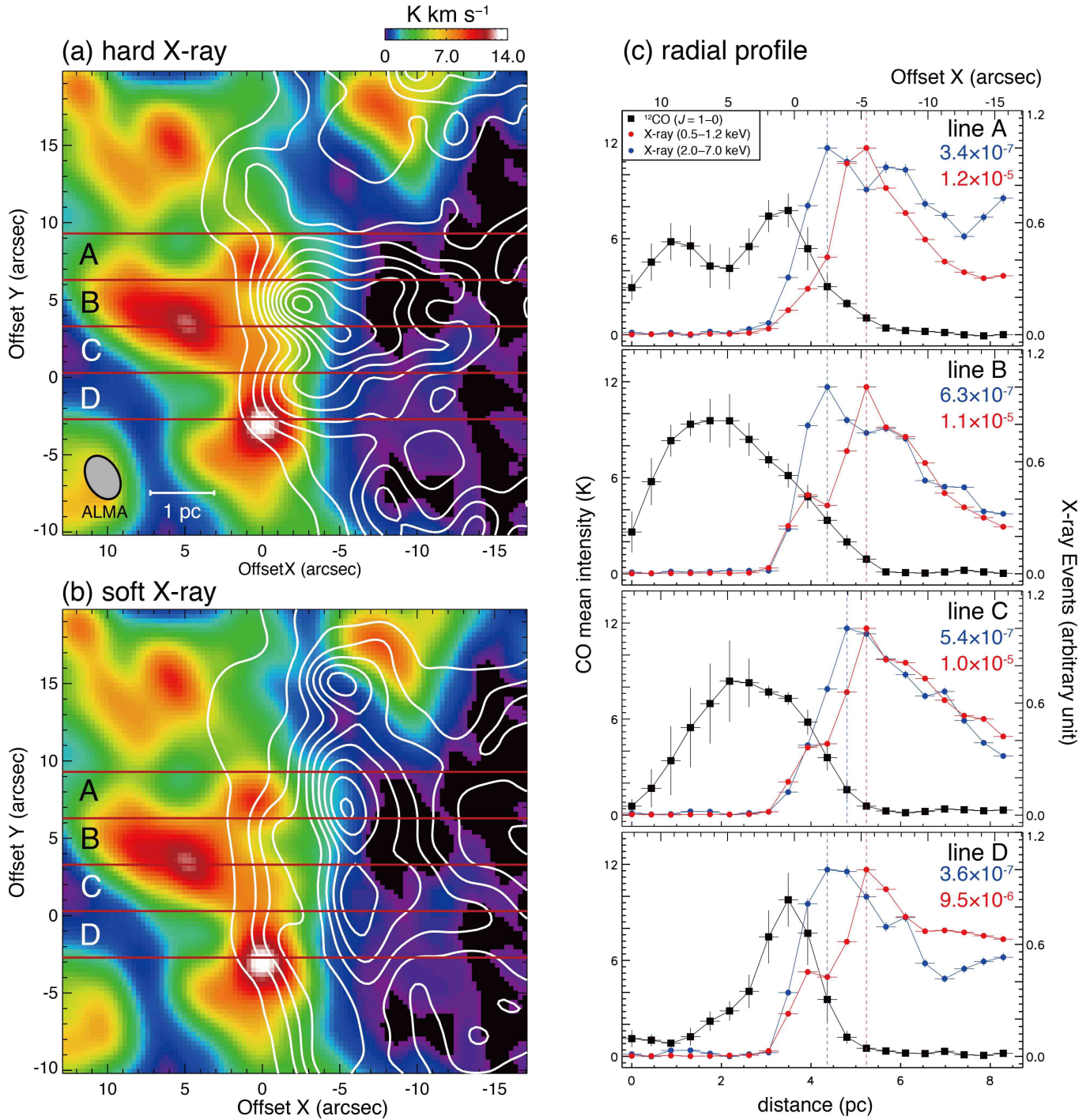}
\caption{(a--b) Enlarged views toward solid regions in Figure \ref{fig7}(b). White contours in (a) and (b) are the same as in Figures \ref{fig4}(a) and \ref{fig4}(b), respectively. (c) Radial Profiles of the $^{12}$CO$(J$ = 1--0) line emission (black), hard X-rays (red), and soft X-rays (blue) for each line as shown in (a) and (b). Red and blue dashed lines indicate the intensity peaks of soft X-rays and hard X-rays, respectively. The values of each panel in blue and red are peak intensities of hard X-ray and soft X-ray, respectively. These units are photons pixel$^{-1}$ s$^{-1}$.}
\label{fig10}
\end{center}
\end{figure*}%


\begin{thebibliography}{99}
\bibitem[Alves(2004)]{2004NewAR..48..659A} Alves, D.~R.\ 2004, \nar, 48, 659
\bibitem[Badenes et al.(2009)]{2009ApJ...700..727B} Badenes, C., Harris, J., Zaritsky, D., \& Prieto, J.~L.\ 2009, \apj, 700, 727
\bibitem[Banas et al.(1997)]{1997ApJ...480..607B} Banas, K.~R., Hughes, J.~P., Bronfman, L., \& Nyman, L.-{\AA}.\ 1997, \apj, 480, 607 
\bibitem[Bilikova et al.(2007)]{2007AJ....134.2308B} Bilikova, J., Williams, R.~N.~M., Chu, Y.-H., Gruendl, R.~A., \& Lundgren, B.~F.\ 2007, \aj, 134, 2308 
\bibitem[Book et al.(2008)]{2008ApJS..175..165B} Book, L.~G., Chu, Y.-H., \& Gruendl, R.~A.\ 2008, \apjs, 175, 165 
\bibitem[Bozzetto et al.(2017)]{2017ApJS..230....2B} Bozzetto, L.~M., Filipovi{\'c}, M.~D., Vukoti{\'c}, B., et al.\ 2017, \apjs, 230, 2 
\bibitem[Cline et al.(1982)]{1982ApJ...255L..45C} Cline, T.~L., Desai, U.~D., Teegarden, B.~J., et al.\ 1982, \apjl, 255, L45 
\bibitem[Cornwell(2008)]{2008ISTSP...2..793C} Cornwell, T.~J.\ 2008, IEEE Journal of Selected Topics in Signal Processing, 2, 793 
\bibitem[Dickel \& Milne(1998)]{1998AJ....115.1057D} Dickel, J.~R., \& Milne, D.~K.\ 1998, \aj, 115, 1057 
\bibitem[Dickel et al.(1995)]{1995ApJ...448..623D} Dickel, J.~R., Chu, Y.-H., Gelino, C., et al.\ 1995, \apj, 448, 623 
\bibitem[Ezawa et al.(2004)]{2004SPIE.5489..763E} Ezawa, H., Kawabe, R., Kohno, K., \& Yamamoto, S.\ 2004, \procspie, 5489, 763 
\bibitem[Feast(1999)]{1999PASP..111..775F} Feast, M.\ 1999, \pasp, 111, 775 
\bibitem[Fruscione et al.(2006)]{2006SPIE.6270E..1VF} Fruscione, A., McDowell, J.~C., Allen, G.~E., et al.\ 2006, \procspie, 6270, 62701V 
\bibitem[Fukuda et al.(2014)]{2014ApJ...788...94F} Fukuda, T., Yoshiike, S., Sano, H., et al.\ 2014, \apj, 788, 94 
\bibitem[Fukui et al.(2003)]{2003PASJ...55L..61F} Fukui, Y., Moriguchi, Y., Tamura, K., et al.\ 2003, \pasj, 55, L61
\bibitem[Fukui et al.(2008)]{2008ApJS..178...56F} Fukui, Y., Kawamura, A., Minamidani, T., et al.\ 2008, \apjs, 178, 56
\bibitem[Fukui et al.(2012)]{2012ApJ...746...82F} Fukui, Y., Sano, H., Sato, J., et al.\ 2012, \apj, 746, 82 
\bibitem[Fukui et al.(2017)]{2017ApJ...850...71F} Fukui, Y., Sano, H., Sato, J., et al.\ 2017, \apj, 850, 71 
\bibitem[Gaensler et al.(2001)]{2001ApJ...559..963G} Gaensler, B.~M., Slane, P.~O., Gotthelf, E.~V., \& Vasisht, G.\ 2001, \apj, 559, 963 
\bibitem[Giacalone \& Jokipii(2007)]{2007ApJ...663L..41G} {Giacalone, J., \& Jokipii, J.~R.\ 2007, \apjl, 663, L41 }
\bibitem[Guo et al.(2012)]{2012ApJ...747...98G} {Guo, F., Li, S., Li, H., et al.\ 2012, \apj, 747, 98 }
\bibitem[Henize(1956)]{1956ApJS....2..315H} Henize, K.~G.\ 1956, \apjs, 2, 315 
\bibitem[Hughes et al.(1998)]{1998ApJ...505..732H} Hughes, J.~P., Hayashi, I., \& Koyama, K.\ 1998, \apj, 505, 732 
\bibitem[Inoue et al.(2008)]{2008stt..conf..281I} Inoue, H., Muraoka, K., Sakai, T., et al.\ 2008, Ninteenth International Symposium on Space Terahertz Technology, 281 
\bibitem[Inoue et al.(2009)]{2009ApJ...695..825I} {Inoue, T., Yamazaki, R., \& Inutsuka, S.-i.\ 2009, \apj, 695, 825}
\bibitem[Inoue et al.(2012)]{2012ApJ...744...71I} Inoue, T., Yamazaki, R., Inutsuka, S.-i., \& Fukui, Y.\ 2012, \apj, 744, 71 
\bibitem[Kawasaki et al.(2002)]{2002ApJ...572..897K} Kawasaki, M.~T., Ozaki, M., Nagase, F., et al.\ 2002, \apj, 572, 897 
\bibitem[Kim et al.(1999)]{1999AJ....118.2797K} Kim, S., Dopita, M.~A., Staveley-Smith, L., \& Bessell, M.~S.\ 1999, \aj, 118, 2797 
\bibitem[Klose et al.(2004)]{2004ApJ...609L..13K} Klose, S., Henden, A.~A., Geppert, U., et al.\ 2004, \apjl, 609, L13 
\bibitem[Kulkarni et al.(2003)]{2003ApJ...585..948K} Kulkarni, S.~R., Kaplan, D.~L., Marshall, H.~L., et al.\ 2003, \apj, 585, 948 
\bibitem[Ladd et al.(2005)]{2005PASA...22...62L} Ladd, N., Purcell, C., Wong, T., \& Robertson, S.\ 2005, \pasa, 22, 62
\bibitem[Laki{\'c}evi{\'c} et al.(2015)]{2015ApJ...799...50L} Laki{\'c}evi{\'c}, M., van Loon, J.~T., Meixner, M., et al.\ 2015, \apj, 799, 50 
\bibitem[Longair(1994)]{1994hea2.book.....L} Longair, M.~S.\ 1994, High energy astrophysics.~Volume 2.~Stars, the Galaxy and the interstellar medium, by Longair, M.~S..~ Cambridge University Press, Cambridge (UK), 1994, 410 p., ISBN 0-521-43439-4  
\bibitem[Maggi et al.(2016)]{2016A&A...585A.162M} Maggi, P., Haberl, F., Kavanagh, P.~J., et al.\ 2016, \aap, 585, A162 
\bibitem[Marsden et al.(1996)]{1996ApJ...470..513M} Marsden, D., Rothschild, R.~E., Lingenfelter, R.~E., \& Puetter, R.~C.\ 1996, \apj, 470, 513 
\bibitem[Matsumura et al.(2017a)]{2017PASJ...69...30M} Matsumura, H., Uchida, H., Tanaka, T., et al.\ 2017a, \pasj, 69, 30 
\bibitem[Matsumura et al.(2017b)]{2017ApJ...851...73M} Matsumura, H., Tanaka, T., Uchida, H., Okon, H., \& Tsuru, T.~G.\ 2017b, \apj, 851, 73 
\bibitem[McMullin et al.(2007)]{2007ASPC..376..127M} McMullin, J.~P., Waters, B., Schiebel, D., Young, W., \& Golap, K.\ 2007, Astronomical Data Analysis Software and Systems XVI, 376, 127 
\bibitem[Minamidani et al.(2011)]{2011AJ....141...73M} Minamidani, T., Tanaka, T., Mizuno, Y., et al.\ 2011, \aj, 141, 73
\bibitem[Naito \& Takahara(1994)]{1994JPhG...20..477N} {Naito, T., \& Takahara, F.\ 1994, Journal of Physics G Nuclear Physics, 20, 477}
\bibitem[Okon et al.(2018)]{2018PASJ..tmp...29O} {Okon, H., Uchida, H., Tanaka, T., Matsumura, H., \& Tsuru, T.~G.\ 2018, \pasj,  }
\bibitem[Otsuka et al.(2010)]{2010A&A...518L.139O} Otsuka, M., van Loon, J.~T., Long, K.~S., et al.\ 2010, \aap, 518, L139 
\bibitem[Park et al.(2003)]{2003ApJ...586..210P} Park, S., Burrows, D.~N., Garmire, G.~P., et al.\ 2003, \apj, 586, 210 
\bibitem[Park et al.(2012)]{2012ApJ...748..117P} Park, S., Hughes, J.~P., Slane, P.~O., et al.\ 2012, \apj, 748, 117 
\bibitem[Pietrzy{\'n}ski et al.(2013)]{2013Natur.495...76P} Pietrzy{\'n}ski, G., Graczyk, D., Gieren, W., et al.\ 2013, \nat, 495, 76 
\bibitem[Rothschild et al.(1994)]{1994Natur.368..432R} Rothschild, R.~E., Kulkarni, S.~R., \& Lingenfelter, R.~E.\ 1994, \nat, 368, 432
\bibitem[Sankrit et al.(2004)]{2004AJ....128.1615S} Sankrit, R., Blair, W.~P., \& Raymond, J.~C.\ 2004, \aj, 128, 1615 
\bibitem[Sano et al.(2010)]{2010ApJ...724...59S} Sano, H., Sato, J., Horachi, H., et al.\ 2010, \apj, 724, 59 
\bibitem[Sano et al.(2013)]{2013ApJ...778...59S} Sano, H., Tanaka, T., Torii, K., et al.\ 2013, \apj, 778, 59 
\bibitem[Sano et al.(2015)]{2015ApJ...799..175S} Sano, H., Fukuda, T., Yoshiike, S., et al.\ 2015, \apj, 799, 175 
\bibitem[Sano(2016)]{2016scir.book.....S} Sano, H. 2016, Shock-Cloud Interaction in RX J1713.7-3946: Evidence for Cosmic-Ray Acceleration in the Young VHE $\gamma$-ray Supernova Remnant, Springer Thesis Series. ISBN 978-4-431-55636-7. Springer Japan 2017 (1st ed.; Tokyo)
\bibitem[Sano et al.(2017a)]{2017AIPC.1792d0038S} Sano, H., Fujii, K., Yamane, Y., et al.\ 2017a,  6th International Symposium on High Energy Gamma-Ray Astronomy, 1792, 040038 
\bibitem[Sano et al.(2017b)]{2017ApJ...843...61S} Sano, H., Yamane, Y., Voisin, F., et al.\ 2017b, \apj, 843, 61 
\bibitem[Sano et al.(2017c)]{2017JHEAp..15....1S} Sano, H., Reynoso, E.~M., Mitsuishi, I., et al.\ 2017c, Journal of High Energy Astrophysics, 15, 1 
\bibitem[Sano et al.(2012)]{2012ApJ...758..126S} {Sano, T., Nishihara, K., Matsuoka, C., \& Inoue, T.\ 2012, \apj, 758, 126 }
\bibitem[Sault et al.(1995)]{1995ASPC...77..433S} Sault, R.~J., Teuben, P.~J., \& Wright, M.~C.~H.\ 1995, Astronomical Data Analysis Software and Systems IV, 77, 433 
\bibitem[Shull et al.(1985)]{1985MNRAS.212..799S} Shull, P., Jr., Dyson, J.~E., Kahn, F.~D., \& West, K.~A.\ 1985, \mnras, 212, 799 
\bibitem[Slavin et al.(2017)]{2017ApJ...846...77S} Slavin, J.~D., Smith, R.~K., Foster, A., et al.\ 2017, \apj, 846, 77 
\bibitem[Sorai et al.(2000)]{2000SPIE.4015...86S} Sorai, K., Sunada, K., Okumura, S.~K., et al.\ 2000, \procspie, 4015, 86
\bibitem[Subramanian \& Subramaniam(2010)]{2010A&A...520A..24S} Subramanian, S., \& Subramaniam, A.\ 2010, \aap, 520, A24 
\bibitem[Torii et al.(2015)]{2015ApJ...806....7T} Torii, K., Hasegawa, K., Hattori, Y., et al.\ 2015, \apj, 806, 7 
\bibitem[Torii et al.(2017)]{2017ApJ...835..142T} Torii, K., Hattori, Y., Hasegawa, K., et al.\ 2017, \apj, 835, 142 
\bibitem[Uchida et al.(2015)]{2015ApJ...808...77U} Uchida, H., Koyama, K., \& Yamaguchi, H.\ 2015, \apj, 808, 77 
\bibitem[Uchiyama et al.(2007)]{2007Natur.449..576U} {Uchiyama, Y., Aharonian, F.~A., Tanaka, T., Takahashi, T., \& Maeda, Y.\ 2007, \nat, 449, 576}
\bibitem[van Loon et al.(2010)]{2010AJ....139...68V} van Loon, J.~T., Oliveira, J.~M., Gordon, K.~D., et al.\ 2010, \aj, 139, 68 
\bibitem[Vancura et al.(1992)]{1992ApJ...394..158V} Vancura, O., Blair, W.~P., Long, K.~S., \& Raymond, J.~C.\ 1992, \apj, 394, 158 
\bibitem[Wong et al.(2011)]{2011ApJS..197...16W} Wong, T., Hughes, A., Ott, J., et al.\ 2011, \apjs, 197, 16 
\bibitem[Wong et al.(2017)]{2017ApJ...850..139W} Wong, T., Hughes, A., Tokuda, K., et al.\ 2017, \apj, 850, 139
\bibitem[Yamaguchi et al.(2014)]{2014ApJ...785L..27Y} Yamaguchi, H., Badenes, C., Petre, R., et al.\ 2014, \apjl, 785, L27 
\bibitem[Yoshiike et al.(2013)]{2013ApJ...768..179Y} Yoshiike, S., Fukuda, T., Sano, H., et al.\ 2013, \apj, 768, 179 
\end{thebibliography}
\end{document}